\documentclass[12pt]{article}

\usepackage{color}
\usepackage{graphics}
\usepackage{graphicx}
\usepackage{epsfig}
\usepackage{textcomp}
\usepackage[small]{caption} 
\usepackage{amsmath,amssymb}
\usepackage{tabularx}
\usepackage[table,usenames,dvipsnames]{xcolor}
\usepackage{hhline}
\usepackage{multicol}
\usepackage{multirow}
\newcolumntype{R}[1]{>{\raggedleft\let\newline\\\arraybackslash\hspace{0pt}}m{#1}} 
\usepackage{enumitem}
\usepackage[normalem]{ulem}
\usepackage{url} 
\usepackage[official]{eurosym}

\usepackage{siunitx}

\usepackage{xr}
\makeatletter
\newcommand*{\addFileDependency}[1]{
  \typeout{(#1)}
  \@addtofilelist{#1}
  \IfFileExists{#1}{}{\typeout{No file #1.}}
}
\makeatother
\newcommand*{\myexternaldocument}[1]{%
    \externaldocument{#1}%
    \addFileDependency{#1.tex}%
    \addFileDependency{#1.aux}%
}
\myexternaldocument{SI_resp}

\usepackage{nameref}

\usepackage[
backend=biber,
style=chem-acs, 
articletitle=true,
date=year,
giveninits=true, 
maxcitenames=1,
maxbibnames=20,
minbibnames=20,
url=false,
isbn=false,
natbib=true,
useprefix=true, 
backend=biber
]{biblatex}
\AtEveryBibitem{%
    \clearfield{doi}%
    \clearlist{language}%
    \clearfield{number}%
    \clearfield{issue}%
}{}

\usepackage{pifont} 

\usepackage{times}

\definecolor{Gray}{gray}{0.9}
\definecolor{DarkGray}{gray}{0.7}
\definecolor{orange}{RGB}{255,127,0}
\definecolor{AaltoOrange}{RGB}{255,127,0}
\definecolor{Peachpuff}{RGB}{255,218,185}
\definecolor{LightCyan}{rgb}{0.88,1,1} 
\definecolor{LightSkyblue}{RGB}{135,206,250}
\definecolor{LightBlueGray}{RGB}{210,228,249}

\title{Electrostatic Discovery Atomic Force Microscopy} 

\author
{Niko Oinonen,$^{1\ast}$, Chen Xu,$^{1\ast}$, Benjamin Alldritt,$^{1\ast}$ \\
	Filippo Federici Canova,$^{1,2}$ Fedor Urtev,$^{1,3}$ Shuning Cai,$^1$ Ond\v{r}ej Krej\v{c}\'i,$^{1}$ \\
	Juho Kannala,$^{3}$ Peter Liljeroth$^{1\dagger}$ and Adam S. Foster$^{1,4\dagger}$\\
\\
\normalsize{$^{1}$Department of Applied Physics, Aalto University, 00076 Aalto, Helsinki, Finland}\\
\normalsize{$^{2}$Nanolayers Research Computing Ltd, London N12 0HL, United Kingdom}\\
\normalsize{$^{3}$Department of Computer Science, Aalto University, 00076 Aalto, Helsinki, Finland}\\
\normalsize{$^{4}$WPI Nano Life Science Institute (WPI-NanoLSI), Kanazawa University,}\\
\normalsize{Kakuma-machi, Kanazawa 920-1192, Japan}\\
\\
\normalsize{$^\ast$These authors contributed equally.}\\
\normalsize{$^\dagger$To whom correspondence should be addressed; }\\
\normalsize{E-mail: peter.liljeroth@aalto.fi; adam.foster@aalto.fi.}
}

\date{}

\begin{document}

\maketitle 

\baselineskip24pt

\begin{abstract}
While offering high resolution of atomic and electronic structure, Scanning Probe Microscopy techniques have found greater challenges in providing reliable electrostatic characterization at the same scale. In this work, we offer Electrostatic Discovery Atomic Force Microscopy, a machine learning based method which provides immediate maps of the electrostatic potential directly from Atomic Force Microscopy images with functionalized tips. We apply this to characterize the electrostatic properties of a variety of molecular systems and compare directly to reference simulations, demonstrating good agreement. This approach offers reliable atomic scale electrostatic maps on any system with minimal computational overhead. 
\end{abstract}

\newpage

\section*{Introduction}
The electrostatic properties of molecules are dominant in a wide variety of processes and technologies, from catalysis and chemical reactions,  \cite{ciampi_harnessing_2018} to molecular electronics \cite{xiang_molecular-scale_2016} and biological functions. \cite{davis_electrostatics_1990} In general, if we can understand the link between molecular function and electrostatics, it offers powerful tools to control and design functionality with nanoscale precision.  \cite{beck_emerging_2020} At this scale, Scanning Probe Microscopy (SPM) is the characterization technique of choice and Scanning Tunneling Microscopy (STM) has become the engine of local electronic characterization for conducting systems,  \cite{zandvliet_scanning_2009,ervasti_single-_2017} while AFM is a general tool for nanoscale imaging without material restrictions.  \cite{loos_art_2005,dufrene_imaging_2017} In high-resolution studies, AFM has evolved from its origins \cite{giessibl_advances_2003} into a breakthrough technique in studies of molecular systems.  \cite{gross_chemical_2009,pavlicek_generation_2017} This has been driven by the use of functionalized tips, and AFM in ultra high vacuum (UHV) now offers a window into molecular structure on surfaces -- aside from the detailed resolution of the results of molecular assembly, it is possible to study bond order, charge distributions and the individual steps of on-surface chemical reactions.  \cite{pavlicek_generation_2017} More recently, additions to the SPM family such as alternate-charging STM \cite{patera_mapping_2019} and single-electron transfer AFM \cite{fatayer_molecular_2019} have offered approaches to study charge behaviour in molecular systems. 

While all these methods give indirect information on the electrostatic properties of the system being studied, significant efforts have been made to develop systematic techniques to directly characterize electrostatic properties. In particular, Kelvin Probe Microscopy (KPFM) was introduced \cite{nonnenmacher_kelvin_1991,sadewasser_kelvin_2018} to simultaneously explore the topography and local contact potential difference with atomic resolution. Despite success in characterizing the electrostatic properties of surfaces,  \cite{palermo_electronic_2006,sadewasser_new_2009,gross_investigating_2014,schulz_elemental_2018} and even proteins,  \cite{sinensky_label-free_2007} the technique has limitations that prevent widespread adoption. Generally, it is experimentally challenging, requiring much longer measurement times than equivalent STM or AFM experiments and can be prone to tip-convolutions.  \cite{zerweck_accuracy_2005} More generally, the varying contributions to the signal mean it is very challenging to obtain quantitative measurements from KPFM.  \citep{songen_weight_2016} The step-change in molecular characterization offered by functionalized tips in AFM has also been harnessed for electrostatic analysis, with KPFM being applied with functionalized tips to provide a local potential maps of single molecules. \cite{mohn_imaging_2012,schuler_contrast_2014} However, the outstanding challenges of KPFM remain, or are even exaggerated: there is no rigorous KPFM theory at the atomic scale, the usually assumed qualitative proportionality to the out-of-plane electric field gradient breaks down at small tip-sample separations  and convolution with unknown background force contributions further complicates the analysis. Attempts to address these problems led to the recent development of scanning quantum dot microscopy,  \cite{wagner_scanning_2015-1} which offers an alternative approach to map the local potential of a surface and adsorbates in high resolution.  \citep{wagner_quantitative_2019} While powerful, the technique relies on quantum dot tip functionalization and a dedicated controller, again limiting its broad implementation as yet. 

In this work, we were inspired by earlier efforts which use AFM tips functionalized with molecules of different electrostatic character to resolve the electrostatic potential by comparing the characteristic distortions of each tip.  \cite{hapala_mapping_2016} However, the limitations of the approach meant it was unable to provide quantitative accuracy. Here, we offer Electrostatic Discovery Atomic Force Microscopy (ED-AFM), a machine learning (ML) approach that can predict accurate electrostatic fields directly from a set of standard experimental AFM images. This methodology offers convenient access to the electrostatic potential of molecules adsorbed on surfaces, which will be important in - for example - understanding their catalytic activity, identifying products of on-surface synthesis routes and facilitating chemical identification of functional groups in unknown molecules.  

\section*{Results and Discussion}
At the heart of the ED-AFM methodology lies a convolutional neural network that is trained to connect input data -- a set of constant height AFM images of the frequency shift $\Delta f$ at different tip-sample distances acquired with two different tip charges -- to a map of the electrostatic potential over the molecule (ES map descriptor). The details of the procedure are given in the Methods section.
\subsection*{Benchmark systems}
\begin{figure*}[!ht]
\centering
\includegraphics[width=140mm]{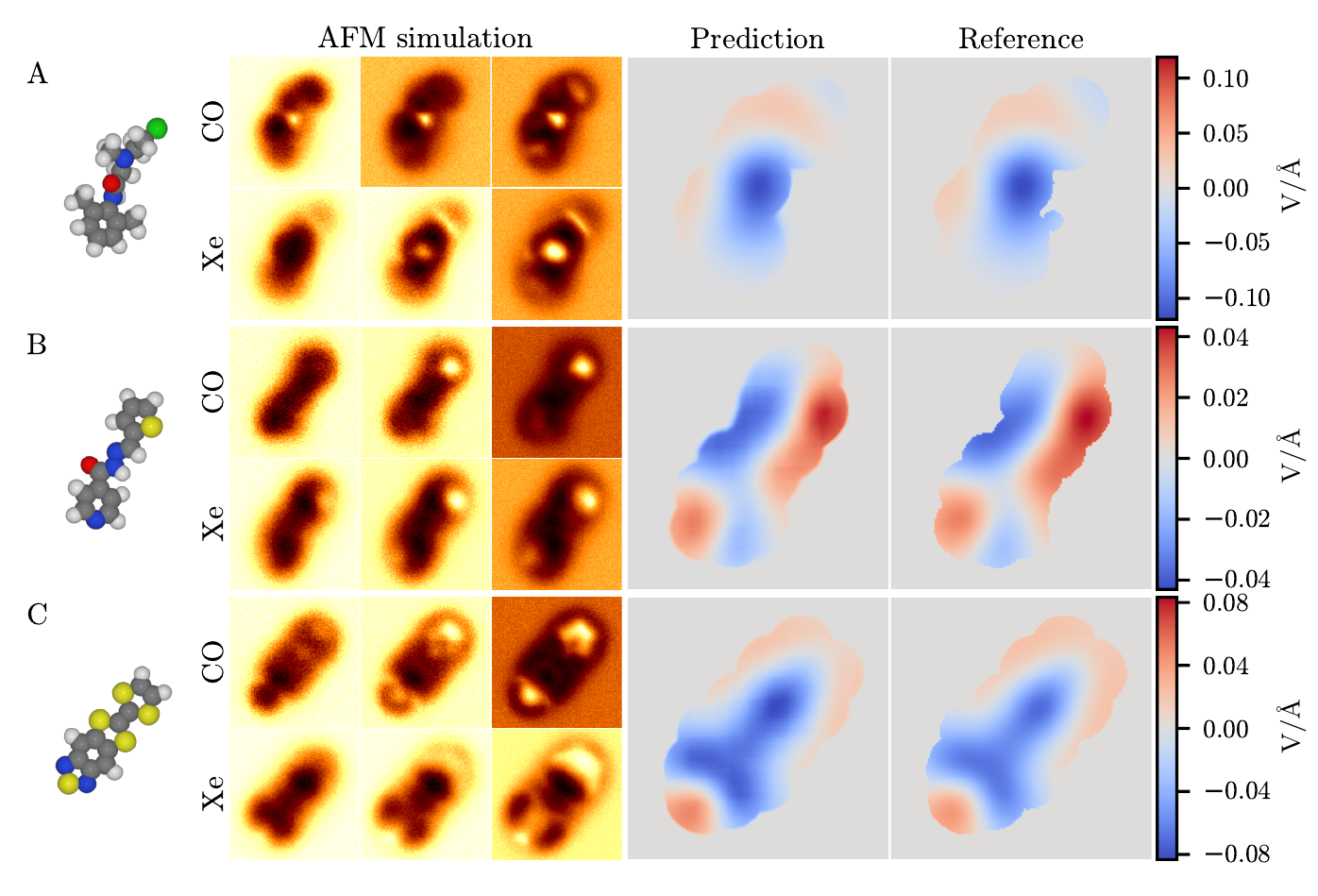}
\caption{Predictions on simulated AFM images. Predictions are shown for three test systems, ({\bf A}) N2-(2-Chloroethyl)-N-(2,6-dimethylphenyl)-N2-methylglycinamide, ({\bf B}) 2-[(1E)-2-thienyl\-methylene]\-hydrazide, and ({\bf C}) tetrathiafulvalene thiadiazole. In each case are shown, from left to right, the 3D structure of the molecule, three out of six input AFM images at different tip-sample distances for both tip functionalizations, and the predicted and reference ES Map descriptors. The colorbar scale for the prediction and the reference is the same on each row.}
\label{sim_full}
\end{figure*}

In order to benchmark the ED-AFM method, we first consider three molecular systems using only simulated data. The first two, "N2-(2-Chloroethyl)-N-(2,6-dimethylphenyl)-N2-methyl\-glycinamide (NCM)" and "4-Pyridinecarboxylic acid, 2-[(1E)-2-thienylmethylene]hydrazide (PTH)" (see Figs.~\ref{sim_full}A,B) were chosen due to the presence of different functional groups and bonds, as well as due to their non-planar geometry. In the third example, we focus on a charge transfer complex, polar tetrathiafulvalene thiadiazole (TTF-TDZ) (see Fig.~\ref{sim_full}C), which was characterized electrostatically in a previous work using KPFM and \emph{ab initio} simulations.  \cite{schuler_contrast_2014} These examples are only considered as free-standing molecules, and therefore the presented orientations and geometries are likely not fully representative of those that the molecules would adopt on a surface.

We consider here the predictions in comparison to the point-charge reference that the model was trained to reproduce. In all cases the match between the predicted and the reference descriptors is generally excellent. The positive and negative regions are predicted in the correct places at correct magnitude with some small imprecision at the edge regions. For a more quantitative comparison, we consider a relative error metric
\begin{equation*}
    \frac{\sum_i \left|y_i-\tilde{y}_i\right|/N}{\max \tilde{y} - \min \tilde{y}},
\end{equation*}
where $y$ is the predicted descriptor, $\tilde{y}$ is the reference descriptor, and the sum is over the $N$ pixels in the descriptor. This is the mean absolute error in the prediction relative to the range of values in the reference. For our three benchmark examples, we find that the relative errors are $1.04\%$ for NCM, $1.79\%$ for PTH, and $2.29\%$ for TTF-TDZ.

\subsection*{Validation}

\begin{figure*}[th]
\centering
\includegraphics[width=160mm]{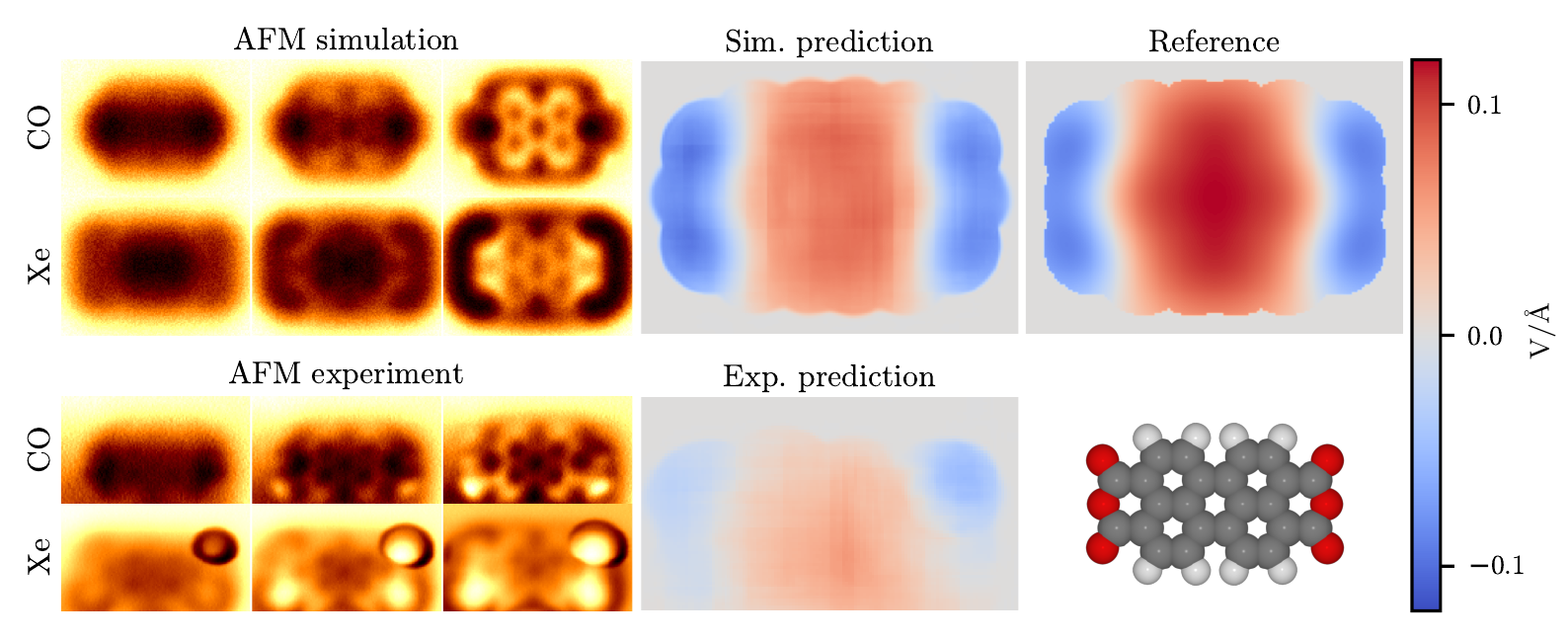}
\caption{Comparison of simulated and experimental predictions for perylenetetracarboxylic dianhydride. On the left are shown three out of six input AFM images at different tip-sample distances for both tip functionalizations, and on the right are the model predictions for both simulation and experiment and the reference descriptor. Both predictions and the reference are on the same colorbar scale. The molecule geometry used in the simulation is shown on the bottom right.}
\label{PTCDA}
\end{figure*}

Clearly, the real test of ED-AFM is with experimental data and we consider three representative example molecular systems. In our first example, we consider perylenetetracarboxylic dianhydride (PTCDA) on the Cu(111) surface (see Fig.~\ref{PTCDA}), a benchmark system in the analysis of molecules on metal surfaces and in characterizing electrostatic interface properties.  \cite{tautz_structure_2007,burke_determination_2009,wagner_quantitative_2019} The simulation is done on a free-standing molecule with planar geometry using point-charges for electrostatics. In the reference ES Map descriptor we find that the ends of the molecule with the three oxygens has a negative field, as would be expected by the electronegativity of oxygen, and the field in the middle of the molecule is positive. In line with the previous simulation examples, the prediction based on the simulated AFM images is in good agreement with the reference. The prediction from the experimental AFM images similarly shows negative field over the ends of the molecule and positive field in between. This matches well with the reference, except for the somewhat weaker magnitude of the field in the experimental prediction.

We note that in the experimental Xe-AFM images, there appears an abnormally large bright feature in the upper right corner over one of the oxygens. The origin of this artifact is unclear, but we speculate that it could be due to a difference in charge at that site (see SI Sec.\ "Possible extra electron in PTCDA" for further discussion). Despite this unusual feature in the AFM images, the model prediction does not appear to be greatly disturbed over the corresponding region. Another unusual feature in this set of images is that there is a gradient in the background of the AFM images decreasing from the upper edge towards the lower edge of the images. This is due to the experiment being performed at a slight tilt with respect to the surface. Originally our model could not handle this feature in the images very well, since such features originating from the surface are normally absent from the simulations that only consider free molecules. After adding artificial gradients in random directions to the simulation images in training, we found that the model begun to perform more consistently on the PTCDA experiment. See SI Sec.\ "Surface tilt effect on model predictions" for further discussion.

\begin{figure*}[ht]
\centering
\includegraphics[width=160mm]{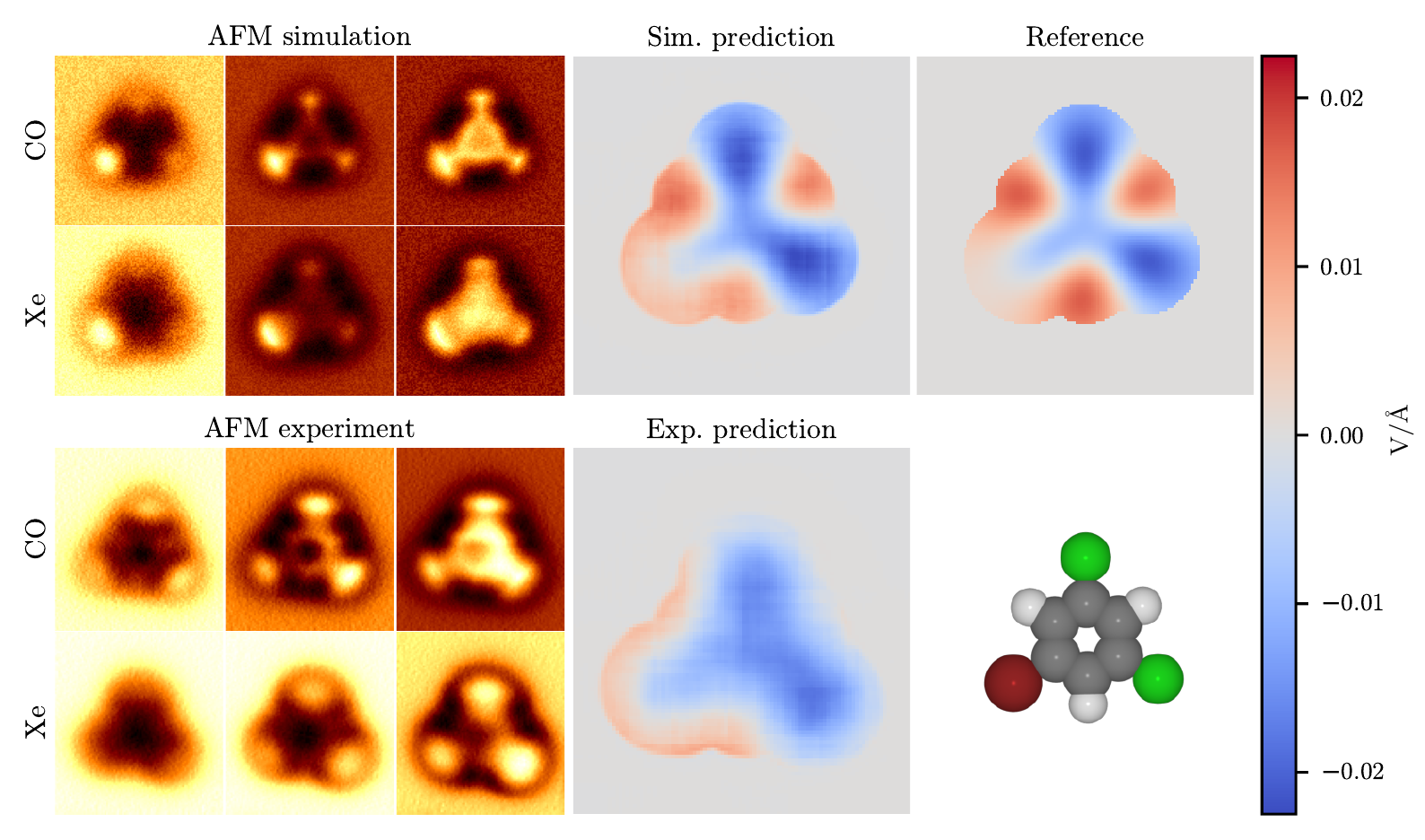}
\caption{Comparison of simulated and experimental predictions for 1-bromo-3,5-dichlorobenzene. On the left are shown three out of six input AFM images at different tip-sample distances for both tip functionalizations, and on the right are the model predictions for both simulation and experiment and the reference descriptor. Both predictions and the reference are on the same colorbar scale. The molecule geometry used in the simulation is shown on the bottom right.}
\label{BCB}
\end{figure*}

For our second example, we study 1-bromo-3,5-dichlorobenzene (BCB) on the Cu(111) surface (see Fig. \ref{BCB}), a planar molecule with mixed halide functionalization. As with PTCDA, for simulations of this molecule we consider a completely planar free-standing molecule with point-charge electrostatics. In the reference ES Map descriptor we find a negative field both over the chlorines and in the middle of the molecule, close to neutral over the bromine, and positive over the hydrogens. Again, the simulation prediction is in good agreement with the reference. In the prediction for the experimental AFM images, we see a region of positive field running along the edge of the lower left part of the molecule, where we suppose the bromine is, and negative field over the other two halides. This matches quite well with the reference, also in the magnitude of the field, except for the missing positive region over the hydrogen opposing the bromine. We note here that the experimental Xe-AFM images have been flipped left-to-right and slightly rotated due to the molecule having rotated between the CO and Xe experiments. The original images can be seen in SI Fig.~S6.

\begin{figure*}[ht]
\centering
\includegraphics[width=160mm]{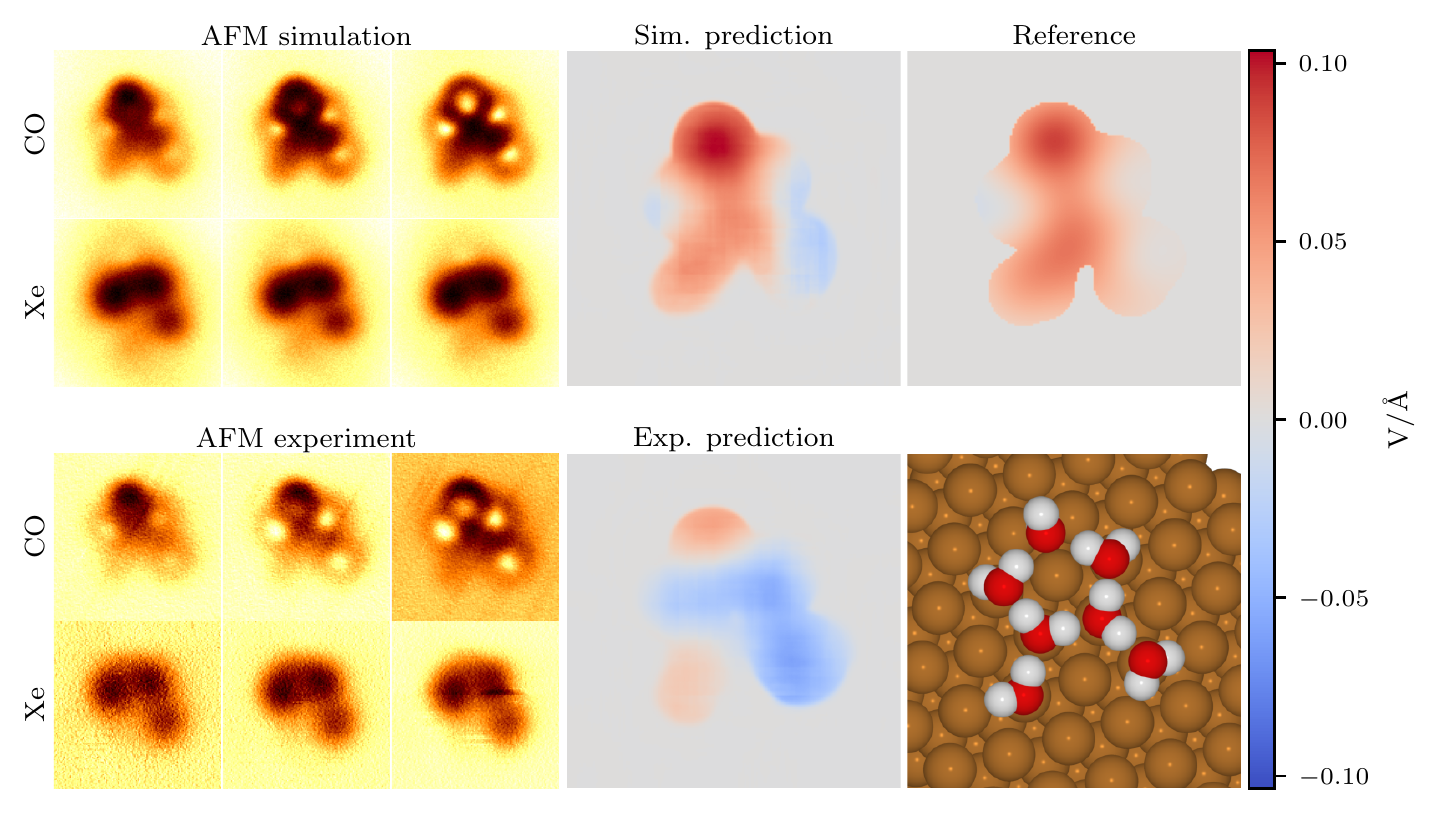}
\caption{Comparison of simulated and experimental predictions for a water cluster on Cu(111). On the left are shown three out of six input AFM images at different tip-sample distances for both tip functionalizations, and on the right are the model predictions for both simulation and experiment and the reference descriptor. Both predictions and the reference are on the same colorbar scale. The molecule geometry used in the simulation is shown on the bottom right.}
\label{water}
\end{figure*}

Our final example is a cluster of seven water molecules on the Cu(111) surface (see Fig. \ref{water}). Again, we use point-charge electrostatics, but this time we include the metallic surface, since this configuration of water molecules is only stable on a surface. The seven water molecules form a single five-member ring with additional two molecules forming an incomplete second ring. Similar water clusters with 9 and 10 molecules forming two complete rings were previously studied with a combination of DFT calculations and STM experiments. \cite{liriano_water_2017} In our calculations, we find that having the second ring be incomplete results in a better match between the simulated and experimental AFM images.

Here in the reference ES Map descriptor we find that the field is mostly positive over the whole structure with some neutral regions on the sides. The simulation prediction matches the reference well except for the sides where the prediction is more negative. The match of the experimental prediction with the reference is reasonable, capturing the positive regions at the top part and at the bottom left "leg" of the structure, but it also has significantly larger negative region in the middle and right side compared to either the simulation prediction or the reference.

Since we have obtained more than the required 6 height-slices for each tip in the experiments, we also consider what happens to the predictions as functions of tip-sample distance for the PTCDA and BCB experiments and find that the predictions stay consistent over small deviations to the distance in either direction (see SI Sec.~"Distance dependence").

\subsection*{Limitations of the current model}

\begin{figure*}[ht]
\centering
\includegraphics[width=150mm]{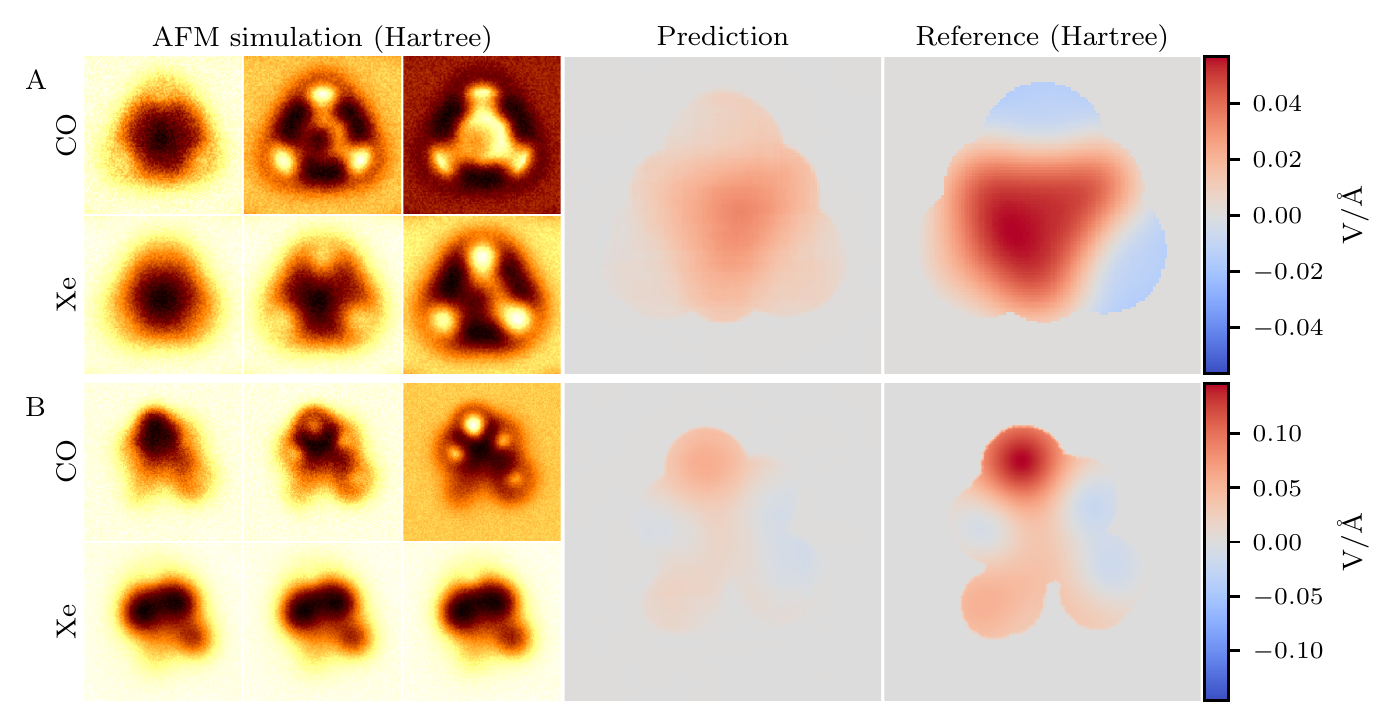}
\caption{Prediction and reference for on-surface geometry of ({\bf A}) BCB and ({\bf B}) water cluster using the DFT Hartree potential for electrostatics in the AFM simulations and for the reference ES Map descriptor.}
\label{BCB_hartree_sim}
\end{figure*}

Until this point we have been considering only the point-charge electrostatics that the ML model was trained on, and found that the model performs well on the simulation examples and the experimental predictions are in fairly good agreement with the ones for simulations. However, the point-charge model of electrostatics has its limitations and in many cases does not perfectly represent the true charge distribution and, by extension, the electric field of the sample. In order to test the validity of the results so far, we performed density functional theory (DFT) calculations using the all-electron density functional theory code FHI-AIMS, \cite{blum_ab_2009} implementing the "tight" basis with the PBE functional \cite{perdew_generalized_1996} and TS van der Waals \cite{tkatchenko_accurate_2009} for all the test systems to obtain their Hartree potentials, which can be used for more accurate electrostatics in both the AFM simulations and the reference ES Map descriptors. We first test the Hartree potentials on our three simulation benchmark systems (Fig.~S7 in SI) and find that the Hartree reference descriptors remain mostly the same as the point-charge ones. The predictions are still qualitatively fairly good, and are at least semi-quantitative in accuracy. The relative errors are $6.12\%$, $4.78\%$, and $7.34\%$ for NCM, PTH, and TTF-TDZ, respectively.

Next, we test the Hartree potential on BCB, one of our experimental test systems. For this test we first relax the geometry on a Cu(111) surface to obtain a more accurate geometry for the molecule. The resulting AFM simulation based on the Hartree potential rather than point charges, the ML prediction, and the reference descriptor are shown in Fig.~\ref{BCB_hartree_sim}A. In the on-surface geometry the bromine is attracted closer to the surface, giving the molecule a slight tilt. This makes the bromine appear less bright in the AFM simulation compared to the planar geometry with point charges, which matches better with the experimental AFM images. The Hartree reference ES Map has similar pattern of field as the point-charge reference over the edges of the molecule, but they differ in the middle over the carbon ring where the Hartree reference has strong positive field instead of negative field. The prediction on the simulation using the Hartree potential correctly catches this positive field in the middle of the molecule, but it misses the negative field over the chlorines and the magnitude of the field is overall too small, roughly by a factor of two. However, the prediction on the experimental AFM images compares more favourably with the ES Map based on point-charge than with DFT/Hartree, even though the Hartree potential from DFT should represent a more accurate description of reality. Note here that it is not \textit{a priori} clear which (if either) ES map should the experiment reproduce: using point charges and the geometry of an isolated molecule is not expected to be the best description of the real BCB molecule on Cu(111) surface. On the other hand, if we run the AFM simulations based on DFT/Hartree molecular structure and electrostatics, we would not expect the ML to reproduce the ES map reference as the model is trained on point charges. Understanding this in detail requires development of a model trained on out-of-distribution samples using the Hartree potential, a focus of future work.

We also perform the DFT calculation for on-surface PTCDA, and find that in this case the model performs poorly even on the simulated data (Fig.~S8 in SI). However, establishing accurate geometries for PTCDA on metal surfaces is very challenging, \cite{ruiz_density-functional_2016} and indeed, when we perform AFM simulations for our on-surface geometry for PTCDA (SI Fig.~S8) we find that the simulation shows an asymmetric and much larger contrast between the ends and the middle of the molecule than in the experimental images. This indicates that the real geometry of the molecule is more planar and more symmetric than in our DFT simulation. Therefore, we feel the reference ES Map descriptor for PTCDA obtained from the DFT Hartree potential here does not provide a valid point of comparison with the experimental prediction.

Finally, we test the Hartree potential on the water cluster (Fig.~\ref{BCB_hartree_sim}B) using the same geometry as with the point charges. Compared to BCB, the difference between the point charges and the Hartree potential is less pronounced, both in the simulated AFM images and the reference ES Map. The biggest difference is the appearance of a stronger positive region at the top of the structure and more negative field on the sides, which takes the reference, at least qualitatively, closer to the ES Map predicted from the experimental AFM images. The prediction on the simulated images using the Hartree potential is qualitatively good, but is weaker in magnitude, especially at the positive region at the top. The relatively weaker magnitude of the predictions compared to the reference seems to be the general pattern for all of our cases using the Hartree potential.

\section*{Conclusions}

This work demonstrates that ED-AFM offers a method for the rapid prediction of electrostatic properties directly from experimental AFM images. We have shown that these predictions demonstrate quantitative accuracy with minimal computational cost once the machine learning infrastructure is trained. In particular, the comparison between reference electrostatics and ML predictions from simulated AFM data has an error of about 1-2\%. However, it is also clear, as for any ML-based approach, that the method cannot predict what it has not learned and it performs poorly in systems where point charge electrostatics are a poor approximation - even for simulated data, the error increases to 5-7\%. We will address this in future work: as part of the database generation, we have stored full density matrices with higher-order quantum chemical accuracy \cite{Crawford:2007:cc,Parrish:2017:psi4} and we are developing methods for their efficient incorporation into the training process. However, we also note that fixed point charges offer decent accuracy in many systems and are routinely used in molecular modelling. \cite{cisneros_classical_2014,riniker_fixed-charge_2018}
The further limitations of ED-AFM lie mainly in the challenges posed for experiments and, in particular, obtaining images of the same system with two different tips. As we have demonstrated, this is feasible on well-defined metal surfaces, but can pose problems on less standard samples. This can be at least partially alleviated by developing methods for the autonomous functionalization of the tip in AFM (see, \emph{e.g.}, https://github.com/SINGROUP/Auto-CO-AFM). We note that the two different tips do not need to be CO and Xe, and we have shown, at least for simulated data, that other pairs would work effectively for ED-AFM.  Finally, as for any machine learning based approach, the predictions can be confused by unexpected artefacts. We apply a suite of tools to improve robustness (see SI Secs.~"Machine learning" and "AFM simulations") and due to the longer range of electrostatic forces, the predictions are  less dominated by images at close approach and hence less sensitive to tip-induced relaxations.  \cite{alldritt_automated_2020}

Finally we note that ED-AFM also offers the prospects of application beyond the examples considered here, to any system where electrostatic characterization is of interest. For example, expansion to predictions for assembled layers,  \cite{grill_nano-architectures_2007,blunt_random_2008} defects \cite{banhart_structural_2011,setvin_surface_2017} and charge dynamics,  \cite{coropceanu_charge_2007} requires only development of the training data to ensure that it is general enough. The method can also be applied to images of systems where direct simulation is impossible due to size, complexity or lack of information. Combined with developments in the autonomous functionalization of the tip in SPM, \cite{gordon_machine_2020} this promises a future of potential in electrostatics for ED-AFM.

\section*{Methods/Experimental}

In general, the adoption of ML methods into materials analysis has seen rapid recent growth \cite{butler_machine_2018,carleo_machine_2019,himanen_data-driven_2019} and this has been followed by an equivalent growth in its applications to image analysis in SPM.  \cite{jesse_principal_2009,woolley_automated_2011,kalinin_bigdeepsmart_2015,kalinin_big_2016,rashidi_autonomous_2018,gordon_automated_2020,leinen_autonomous_2020,gordon_machine_2020,azuri_role_2021} Here we build upon our ML method for predicting molecular structure from AFM images, \cite{alldritt_automated_2020} to predict the electrostatic field of the sample molecule.

\begin{figure*}[ht]
\centering
\includegraphics[width=160mm]{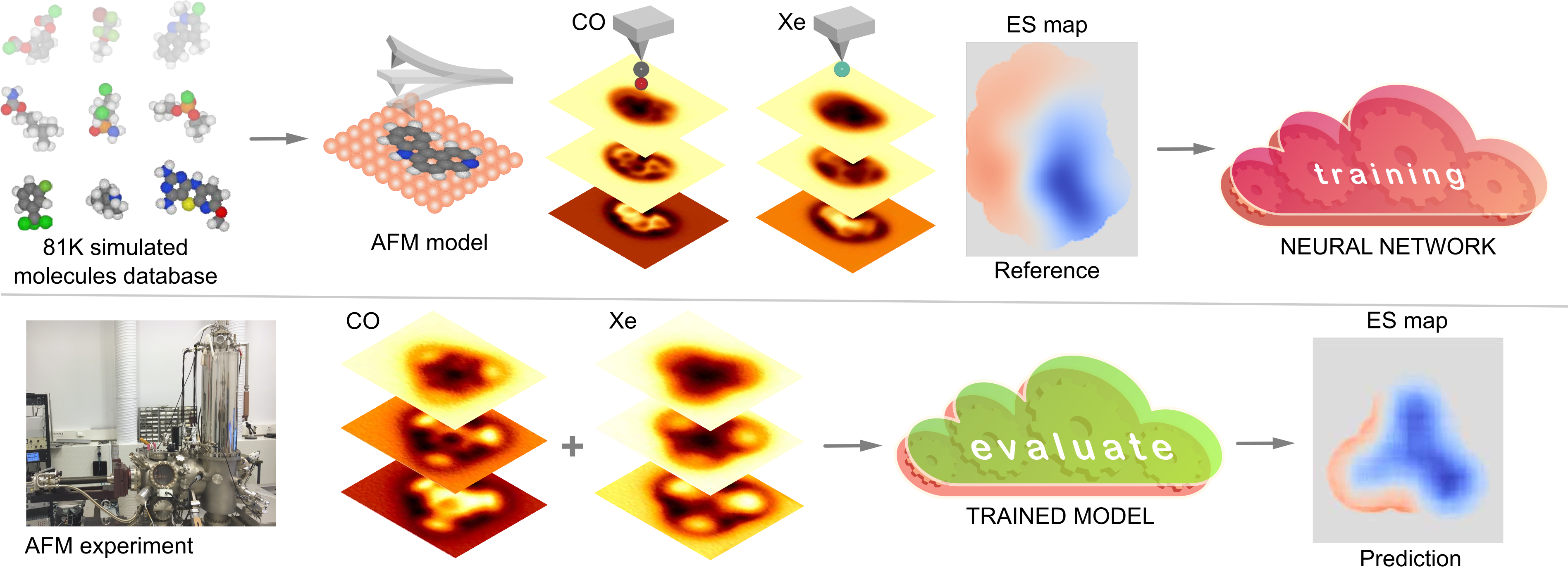}
\caption{Schematic of the ED-AFM method. We train a neural network that takes two sets of AFM images as input and translates them to the ES Map descriptor, which is the vertical component of the electrostatic field over the sample molecule. The model is trained on simulated sets of input-output pairs calculated from a database of several tens of thousands of molecule geometries. The trained model can then be applied to experimental AFM images to produce a prediction of the sample electric field.}
\label{method}
\end{figure*}

The overall idea of our method is illustrated in Fig.~\ref{method}. We train a deep learning model on a dataset of simulated AFM images and reference descriptors based on a large database of molecular geometries, including electrostatics at the level of point charges taken from the associated quantum chemistry calculations. \cite{alldritt_automated_2020} For more details on how the dataset is generated from the geometries, see Sec.\ "Data set" in the Supplementary Information (SI). The trained ML model can then be used to make predictions with experimental data as input. More specifically, the ML model takes as input two sets of six AFM images of the same sample at different tip-sample distances, imaged with two different functionalizations of the AFM tip. The model, which is an Attention-U-Net-type convolutional neural network \cite{ronneberger_u-net_2015,oktay_attention_2018} (more details in SI Sec.\ "Machine learning"), translates the AFM images into a descriptor of the imaged sample, which we call the Electrostatic (ES) Map. The ES Map is defined as the vertical component of the electrostatic field calculated on a constant-height surface \SI{4}{\angstrom} above the highest atom in the sample. Furthermore, the non-zero region is cut only into the region where the sample molecule of interest is visible in the AFM images, such that the ML model is not asked to predict the field over the background where there are no discernible features present. For more detailed description of how the ES Map descriptor is generated, see SI Sec.~"ES map descriptor".

The use of two sets of AFM images in the input is motivated by the observation that the different distortions in the images obtained with different tip functionalizations are linked to the different electronic charges on the tips.  \cite{hapala_mapping_2016} Thus, given a database of such pairs of images, an ML model should be able to learn what role the electrostatics play in the formation of the images and separate the electrostatic contribution from other forces that contribute to the images. Here for the tip functionalizations we use CO, which has a slightly negative charge, and Xe, which has somewhat positive charge. Other choices for functionalization are possible, and we investigate the alternative tip combinations of Cl-CO and Cl-Xe on simulated data in SI Sec.\ "Other tip combinations".  We also tried training a model using images of only a single tip functionalization with CO, but found the results to be less robust (see SI Sec.\ "Single-channel measurements"). As in our previous work,  \cite{alldritt_automated_2020} we train the model using simulated AFM images and validate the model using both simulated and experimental AFM images. Our implementation of the model in Pytorch \cite{paszke_pytorch_2019} with pretrained weights can be found at \url{https://github.com/SINGROUP/ED-AFM}.
\newpage

\subsection*{Supporting Information}
The Supporting Information contains details of the machine learning models and training data, including investigation of the sensitivity of the predictions to tip-surface distance, noise, various simulation parameters and the number of channels used in experiments. It also contains expanded discussion of the AFM simulations and experiments themselves.

\subsection*{Acknowledgements}
Computing resources from the Aalto Science-IT project and CSC, Helsinki are gratefully acknowledged. This research made use of the Aalto Nanomicroscopy Center (Aalto NMC) facilities and was supported by the European Research Council (ERC 2017 AdG no.~788185 “Artificial Designer Materials''), and the Academy of Finland (Projects no. 311012, 314862, 314882, Centres of Excellence Program project no. 284621, and Academy professor funding no.~318995 and 320555). ASF has been supported by the World Premier International Research Center Initiative (WPI), MEXT, Japan. OK and CX acknowledge funding from the European Union’s Horizon 2020 research and innovation programme under the Marie Skłodowska-Curie grant agreements “QMKPFM” No 845060 and "EIM" No 897828.

\printbibliography[]

\end{document}


\maketitle 

\baselineskip24pt

\newpage

\section*{Methods}

\subsection*{Machine learning}\label{sec:ml}

The core of our model has the structure of the U-Net \cite{ronneberger_u-net_2015}, where the feature maps first enter an encoder which down-samples them with pooling layers and then enter a decoder which up-samples them back to the original size, with skip connections between the layers of matching size in the encoder and decoder. In addition to the different number of channels and layers, the main difference to the original U-Net is that we start the network with 3D convolutions and then change to 2D convolutions in the middle, and we use Attention-Gate (AG) \cite{oktay_attention_2018} layers in the skip connections. Around the core, we have the input stage which merges the two input sets of AFM images, and the output stage which outputs the ES Map descriptor. All of the convolutional layers use replicate padding, and, except for the output layer, all convolutional layers are followed by LeakyReLU activations \cite{xu_empirical_2015} with negative slope of $0.1$.

\begin{figure*}[ht]
\centering
\includegraphics[width=150mm]{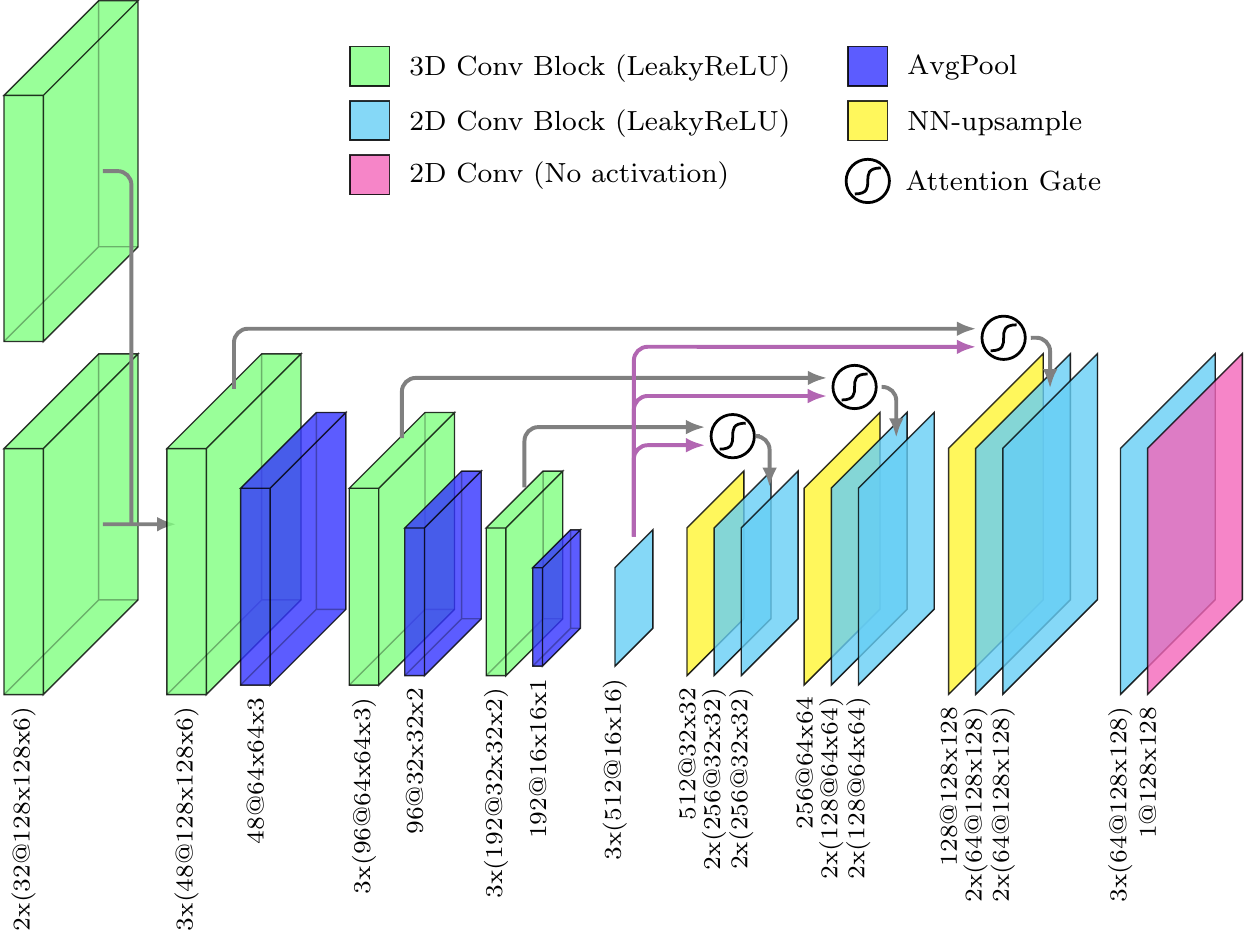}
\caption{Schematic illustration of the model architecture. Below each layer or block of layers, the output shape of the layer is reported in the format (number of channels)@(feature map size) assuming a $128\times128$ lateral input size, and in the blocks the first number indicates the number of layers of that shape in the block. Here, NN = nearest neighbour.}
\label{model_params}
\end{figure*}

The model architecture, along with the shapes of the layers assuming $128\times128$ lateral input size, are illustrated in Fig.~\ref{model_params}. At the start, the two inputs are fed into their own blocks of two 3D convolutions with 32 channels. The outputs from these are concatenated together in the channel dimension and then fed into the encoder. The encoder consists of three blocks of 3D convolutions and poolings. In each block, there are three 3D convolutions, with 48, 96, and 192 channels in each block, respectively, and the poolings are AvgPool layers. The poolings all have pool regions of $2 \times 2 \times 2$, but the middle pooling has a stride of $2 \times 2 \times 1$, so that the size of the feature map in $z$-direction is only reduced by 1. After the last pooling layer, the 3D feature maps are transformed into 2D feature maps by concatenating the remaining $z$-layers of the 3D feature maps into channels of the resulting 2D feature maps. The middle section between the encoder and the decoder has a block of three 2D convolutions with 512 channels. The decoder has three up-sampling stages corresponding to the three down-sampling stages of the encoder. At each stage there is first a nearest-neighbour up-sampling followed by a block of two 2D convolutions. Then the skip connection from the corresponding stage of the encoder is passed through the AG and is concatenated as additional channels to the input of a second block of 2D convolutions. The convolution blocks in the decoder stages have 256, 128, and 64 channels. After the decoder, the model has one more 2D convolution block with three 2D convolutions and 64 channels, and one more 2D convolution with a single channel to output the ES Map descriptor. The total number of parameters in the model is 15,604,900.

\begin{figure*}[ht]
\centering
\includegraphics[width=160mm]{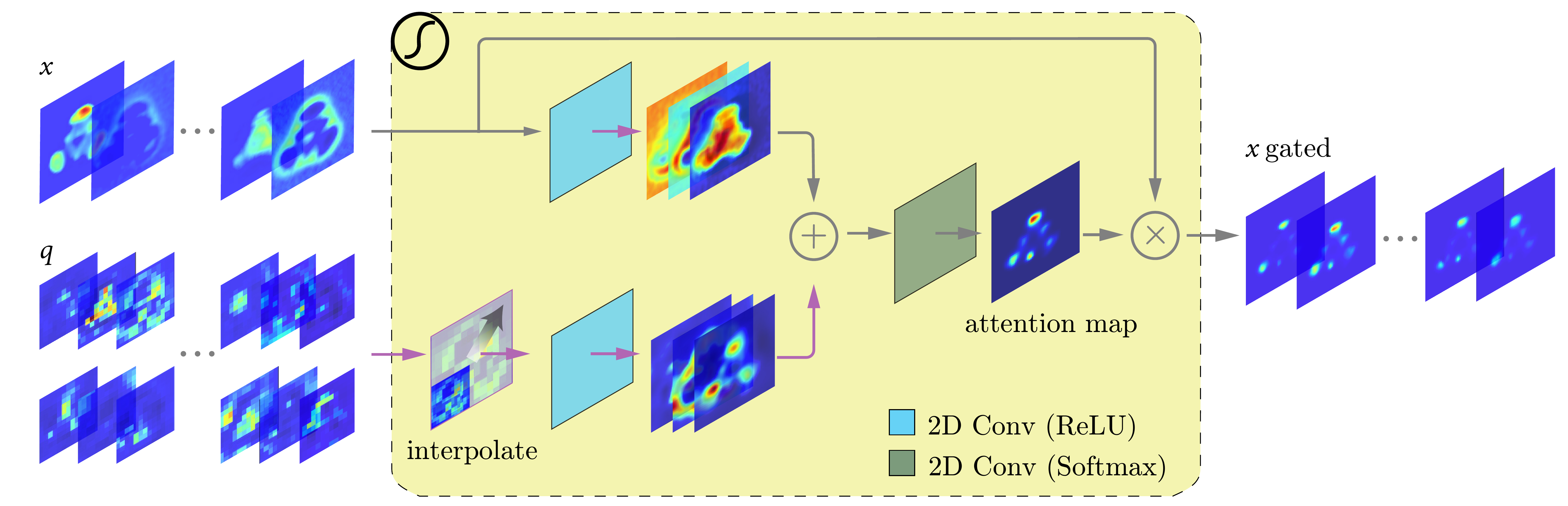}
\caption{Schematic illustration of the Attention Gate (AG) using the BCB molecule as an example. Randomly picked features maps in inputs and outputs are presented. AG operates with 2 inputs: skip connection feature maps (input $x$) together with compressed representation at the end of the encoder (query $q$). Since all three skip connections have different sizes than the query, $q$ is interpolated to match the size of $x$. Both $x$ and $q$ are passed through 2D convolutions with ReLU activation, and then they are summed together and the result is passed through a 2D convolution layer with Softmax activation to produce the attention map. The attention map is finally multiplied pixel-wise with the skip connection features maps to produce the gated output of the AG layer.}
\label{AG_schem}
\end{figure*}

In the proposed architecture we implemented AGs \cite{oktay_attention_2018} on the skip connections from the encoder to the decoder. Based on a specific task, AGs can suppress irrelevant and highlight useful parts in inputs. An AG architecture is illustrated in Fig.~\ref{AG_schem}. It has two inputs: a set of feature maps from 3D convolution blocks in encoder flattened into 2D feature maps ($x$) and a query ($q$) -- feature maps from the last 2D convolution layer in the middle part of the model. For a $128\times128$ lateral input size, the skip connections have the following shapes: 288@128x128, 288@64x64, 384@32x32 and the query shape is 512@16x16. After an interpolation of $q$ to match the shape of $x$, both $x$ and $q$ are passed through independent 2D convolution layers with ReLU activations and then combined together by channel-wise summation. The feature maps are then passed through a 2D convolution with a Softmax activation to create a single-channel map of attention coefficients -- the attention map. Finally, the attention map is mixed with the skip connection by element-wise multiplication in each channel. Due to the construction with a Softmax activation, the AG learns to highlight the most relevant regions in the input without explicitly being trained to do so.

Our objective function is the mean squared error
\begin{equation}
	\mathrm{MSE}(y, \tilde{y}) = \frac{1}{N} \sum_{i=0}^{N-1} (y_i - \tilde{y}_i)^2,
\end{equation}
where $y$ is the predicted ES Map, $\tilde{y}$ is the reference ES Map, and the sum is over the $N$ pixels. For reference, the losses on the final trained model are $2.17 \times 10^{-5}$ on the training set, $2.49 \times 10^{-5}$ on the validation set, and $2.47 \times 10^{-5}$ on the test set. The parameters are optimized using the Adaptive Moment Estimation (Adam) optimizer \cite{kingma_adam_2017}, with learning rate $10^{-4}$ and the default values of $\beta_1 = 0.9$ and $\beta_2 = 0.999$ for the moment decay parameters. Additionally, we use a learning rate decay, where on each iteration $i$, the initial learning rate is multiplied by a factor
\begin{equation}
	\frac{1}{1+10^{-5} \cdot i}.
\end{equation}
The training set has a total of 6000 batches with 30 samples each, and the model is trained for a total of 50 epochs. The dataset is described in more detail below in Sec.\ "\nameref{sec:data_gen}".

During training, we preprocess the samples in several ways before they enter the model. The samples are normalized by subtracting the mean and dividing by the standard deviation per each height-layer in the AFM image stack. For regularization, we randomly add to each sample noise, pixel shifts, cutouts, and additive background gradient planes, and the samples are randomly rotated, flipped, and cropped. The noise augmentation is discussed below in Sec.\ "\nameref{sec:noise}". The pixel shifts are applied independently to each layer in the AFM image stack, such that the pixel values roll over the borders. The maximum shift between adjacent slices in the AFM image stack is 2\% of the image size and maximum total shift is 4\% of the image size. The cutouts randomly erase an area of the input image. The erased area for each cutout is at most 1\% of the total area of the image and has a maximum aspect ratio of 1 to 10. A maximum of five cutouts are added to each image with 20\% probability for each one. For details of the background gradient augmentation, see below Sec.\ "\nameref{sec:surface_tilt}". The original samples generated in a size of $192\times192$ are rotated to a random angle by bicubic interpolation, flipped up-down with $50\%$ probability, and then cropped to size $128\times128$ to get rid of any empty pixels in the corners. The images are then further cropped to a random position at random size of at minimum 75\% of the original size and a random aspect ratio of at most 1.25 in either direction.

The training samples are all generated on a $24 \times 24$ \si{\angstrom^2} frame discretized on a $192 \times 192$ grid, and the molecule is always in the center of the frame. Since the model is trained on this specific pixel density of $\SI{24}{\angstrom} / \SI{192}{\mathrm{pixels}} = \SI{0.125}{\angstrom/\mathrm{pixel}}$, the experimental images are resized to match this resolution before entering the model. Additionally, we always crop the images into multiples of 8 pixels in each dimension in order to keep the dimensions consistent over the pooling and upsampling layers in the model (three halvings = 1/8 image size). The experimental images are also normalized in the same way as the simulated training samples.

\subsection*{Distance randomization}\label{sec:dist_rand}

In AFM experiments, it is often difficult to know the exact distance between the tip and the sample, and the distance range where the tip-sample interaction is stable differs between samples. These facts mean that the range of distances available in AFM images is variable. In order to be robust against varying tip-sample distance, we randomize it during the generation of the training simulation samples within a $\SI{0.5}{\angstrom}$ window. Here we have to take into account the additional factor of the second tip. It is unclear whether the ML model would benefit from having the two tips at the same distance from the sample or if it would also work if the tips are not aligned.

\begin{figure*}[ht]
\centering
\includegraphics[width=160mm]{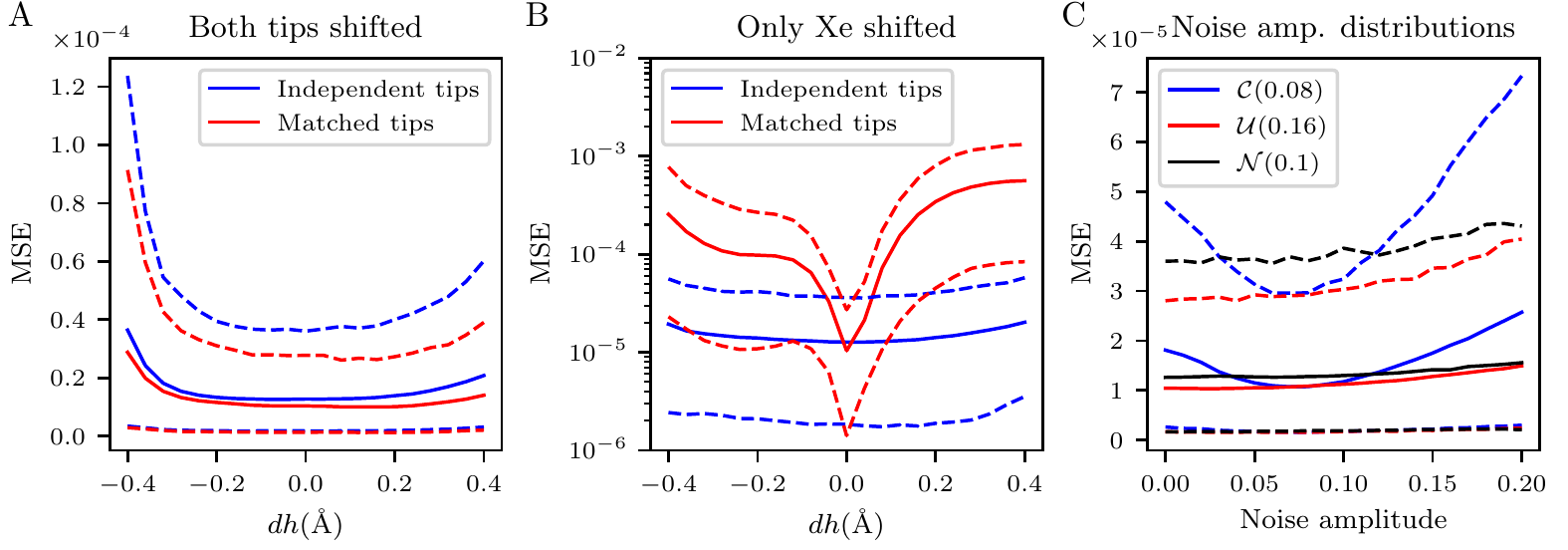}
\caption{Loss statistics with different randomizations of the tip-sample distance and the noise amplitude on a subset of the test set. (A, B) The MSE loss as a function of tip-sample distance offset $dh$ with (A) both tips offset and (B) only Xe offset for two models trained with independently randomized and jointly randomized distance for the two tips. Here $dh = \SI{0}{\angstrom}$ represents the average distance used in the training. (C) The MSE loss as a function of noise amplitude for three different models trained with constant ($\mathcal{C}$), uniform random ($\mathcal{U}$), and normally distributed ($\mathcal{N}$) noise amplitudes. In all plots, the solid lines represent the mean loss, and the dashed lines represent the 5th and 95th percentiles, so that 90\% of the losses are contained within the region enclosed by the dashed lines of the same color.}
\label{stats}
\end{figure*}

In order to test this, we train the ML model with two differently generated datasets, one with matched tip distance where the tip-sample distance is the same for the two tips for the same training sample, and one with independently randomized tips. We then test how the MSE loss behaves for the two differently trained models as a function of the tip-sample distance on a subset of 3000 samples from the test set (Fig.~\ref{stats}A,B). For matched tip distances on the test samples (Fig.~\ref{stats}A), we find that both models have almost a flat loss curve within the window of distances used in the training, but outside of that window the loss starts to increase. The increase in loss is especially sharp on the side of smaller distances. We discuss why too close distances are undesirable at more length in the context of the experimental predictions below in Sec.\ "\nameref{sec:distance}". In these results the difference between the matched and independent tips is small, with possibly a small advantage in favor of the matched tips. However, when we do the test such that the CO-tip is held at constant height and the Xe tip is shifted, the difference in performance becomes very apparent. The model trained on matched tips does well for the zero-shift where the tip distances are matched, but when the Xe distance is varied the loss becomes significantly bigger, by more than a order magnitude even within the training window of distances. This is in contrast with the model trained with independently randomized tips, which has similarly flat loss curve as in the first test. Clearly, any small disadvantage for the independent tip randomization in the first test is worth the trade-off for significantly improved stability when the tip distances are not exactly matched.

\subsection*{Noise amplitude distributions}\label{sec:noise}

Experimental AFM images always have some level of noise present in the values of the pixels. In order to be robust against noise in the input images, we add noise with random uniform distribution to the simulated images during training. Since the noise is independent between training epochs, this also serves as a type of regularizing augmentation that reduces overfitting of the model. The generated noise is multiplied by the range of the values ($\max - \min$) in the sample before the addition operation to keep the level of the noise consistent between the samples. We test here three different ways of choosing the amplitude of the noise: constant amplitude, uniform random amplitude, and normally distributed amplitude. For the normal distribution, we use the absolute value of the generated value as the amplitude, and we choose the standard deviation of the normal distribution to be $0.1$. This gives the noise amplitude an expected value of ${\small \sim} 0.08$. To keep the average level of the noise consistent between the tests, for the constant amplitude we choose the value $0.08$, and for the uniform random amplitude we choose the range $[0, 0.16]$.

We test these three differently trained models on a subset of 3000 samples from the test set. The average MSE loss on these test samples as a function of the noise amplitude is presented in Fig.~\ref{stats}C. The most striking feature here is the difference between the constant amplitude model and the random amplitude models. The model trained with the constant amplitude does the best on the amplitude of the noise that it was trained on and has worse loss for all other amplitudes, including the zero-amplitude without any noise. This is saying that for this model clean images are harder to interpret than noisy ones, clearly an undesired behaviour. In contrast, the models trained with random noise amplitude have much flatter loss curves, with the uniform random amplitude having a small advantage over the normally distributed one. Further tests would be needed to determine what is the optimal distribution for the random noise amplitude, but it is clear that random amplitude for the noise is better than constant amplitude. For the training of the model used for the predictions in the main article, we used the normally distributed amplitude.

\subsection*{Dataset}\label{sec:data_gen}

Our model training is based on a database of 81086 molecular geometries containing the elements H, C, N, O, F, Si, P, S, Cl, and Br. The distribution of the elements in the molecules is shown in Table~\ref{tab:elements}. Here we can see that the distribution is not even: H and C are contained in almost every molecule with N and O being very common as well, but the rest of the elements are significantly less common. In our previous work on molecule structural prediction from AFM images we used a simple criterion with a fixed number of rotations for each molecule to choose the molecule orientations for the samples \cite{alldritt_automated_2020}. This lead to an overemphasis on the more common elements, especially H, in the dataset. Here we have chosen the rotations for the molecules more carefully in order to make the element distribution more even in the dataset.

For choosing the rotations, we want to consider what elements are close to the surface of the molecule, so that those atoms could possibly be seen in the AFM images. To this end, we compute the convex hull of the molecule, yielding us sets of three points that define planes on the surface of the molecule. We consider each one of these planes in turn and include the rotation corresponding to the plane probabilistically based on the elements close to the plane, choosing the probabilities such that the rarer elements are emphasized. An element is considered to be close to the plane if an atom with that element is within \SI{0.7}{\angstrom} of the plane. To counter any bias that using the convex hull planes may incur, we also choose completely random rotations of the molecules, which are again included probabilistically emphasizing the rarer elements. Finally, we noted that the database does not contain many completely flat geometries, so we include any rotations of the molecules that contain a planar segment, which we define as a plane on the surface of the molecule which contains at least 10 atoms within \SI{0.1}{\angstrom} of the plane. In order not to have overlap between the rotations, no rotations within $5\degree$ of each other for the same molecule are included.

Using this procedure, we generate a total of 235554 different orientations of the molecules, which we divide into training, validation, and test sets as 180000, 20000, and 35554 samples, respectively. We take care not to include any of our test molecules in the training or validation sets. The distribution of the elements contained in the final chosen rotations based on the \SI{0.7}{\angstrom} criterion is shown in Table~\ref{tab:elements}. H and C are still the most common elements, and this is natural, since any orientation where one of other elements is seen, likely H and/or C is also there. The occurrence of the rest of the elements is now more even, except for Si and P, which are mostly only contained inside the molecules and therefore are not often seen close to the surface.

\begin{table}[!htb]
    \centering
    \begin{tabular}{c c c}
        Element & \% of molecules in database & \% of chosen rotations \\
        \hline
        H & 99.3 & 87.3 \\
        C & 99.6 & 49.8 \\
        N & 60.8 & 24.8 \\
        O & 76.5 & 29.5 \\
        F & 5.4 & 16.5 \\
        Si & 1.5 & 0.2 \\
        P & 3.1 & 1.2 \\
        S & 15.0 & 23.4 \\
        Cl & 13.7 & 28.6 \\
        Br & 3.1 & 16.4 
    \end{tabular}
    \caption{Distribution of different elements contained in the molecules in our database and in the rotations of the molecules that we chose. For the chosen rotations an element is included in the count if it is contained in the region up to \SI{0.7}{\angstrom} below the top-most atom in the molecule.}
    \label{tab:elements}
\end{table}

\subsection*{ES Map descriptor}\label{sec:esmap}

\begin{figure*}[!ht]
\centering
\includegraphics[width=140mm]{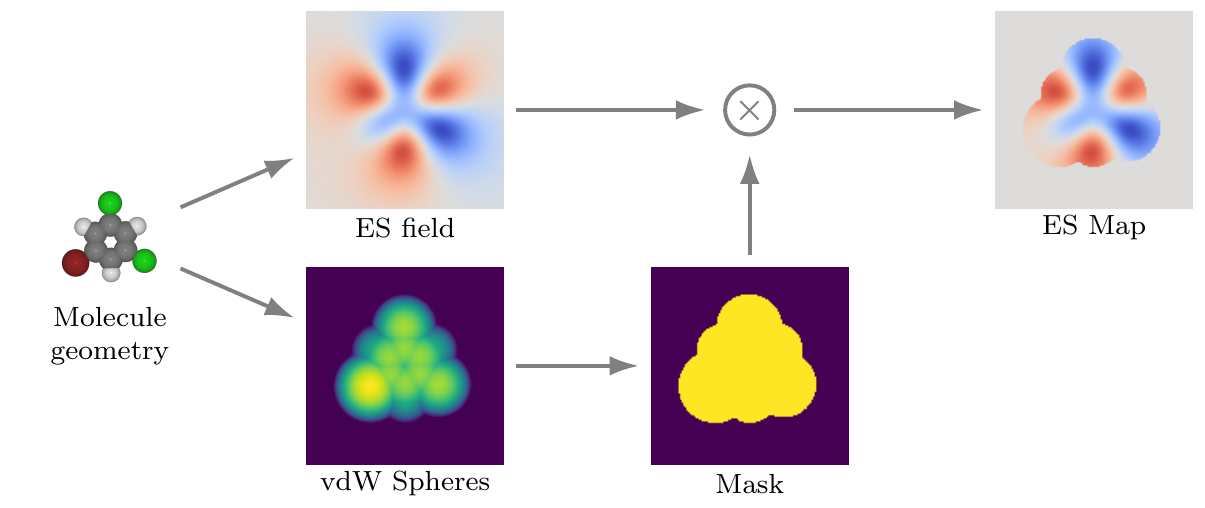}
\caption{Schematic illustration of the process for generating the ES Map descriptor using the BCB molecule as an example. The molecule geometry is used to compute both the ES field and the vdW-spheres descriptor. The vdW-Spheres descriptor gets turned into a binary mask which is then multiplied pixel-wise with the ES field to produce the ES Map descriptor.}
\label{esmap_schem}
\end{figure*}

The ES Map descriptor is the z-component of the ES field originating from the charges in the sample molecule, calculated at a constant-height surface \SI{4}{\angstrom} above the top-most atom in the molecule, and then cut to be non-zero only in the region occupied by the molecule. This process is illustrated in Fig.~\ref{esmap_schem}. We define the z-direction to be parallel to the oscillation direction of the AFM probe, which is perpendicular to the hypothetical surface which the molecule would be sitting on, and the positive z-direction is pointing away from the surface (out of the page in the figures here). We find the highest z-coordinate of a center of an atom in the molecule and then add \SI{4}{\angstrom} to that value to define the z-coordinate of the constant-height surface of the ES Map. The xy-coordinates of the pixels form a grid corresponding to the matching pixel coordinates in the AFM images. Then for a given pixel $\mathbf{R}_{ij} = (x_i, y_j, z)$ the value of the pixel is
\begin{equation}
    E_{\mathrm{z}}(\mathbf{R}_{ij}) = k_{\mathrm{e}} \sum_{k=1}^{n} \frac{q_k (\mathbf{R}_{ij} - \mathbf{r}_k) \cdot \hat{\mathbf{z}}}{|\mathbf{R}_{ij} - \mathbf{r}_k|^3},
\end{equation}
where $k_{\mathrm{e}}$ is the Coulomb constant, $q_k$ is the charge and $\mathbf{r}_k$ is the coordinate of the $k$th atom in the molecule, $n$ is the number of atoms in the molecule, and $\hat{\mathbf{z}}$ is a unit vector in the z-direction. For restricting the non-zero area, we use the vdW-Spheres descriptor, which we introduced in our previous work \cite{alldritt_automated_2020}. We modify the descriptor here by adding a constant \SI{1}{\angstrom} to the vdW radii of the atoms and restrict the deepest coordinate to be \SI{2}{\angstrom} below the top coordinate. We then turn the vdW-Spheres descriptor into a binary mask by setting the background values to 0 and all other values to 1. This mask is then multiplied pixel-wise with the $E_{\mathrm{z}}$-values computed earlier to produce the final pixel values of the ES Map descriptor.

\subsection*{AFM Simulations}\label{sec:simulations}

We simulate AFM images using the probe particle model \cite{hapala_mechanism_2014}. The procedure for generating the training samples is explained in our previous work \cite{alldritt_automated_2020}. Here we additionally do simulations with the Xe and Cl probe-particle tips. The lateral spring constants we use for the probe particles are \SI{0.25}{N/m} for CO and Xe, and \SI{0.5}{N/m} for Cl. The radial spring constant is \SI{30}{N/m} in all cases. The CO and Xe tip charges are modelled as quadrupoles with quadrupole moments of \SI{-0.1}{e \times \angstrom^2} and \SI{0.3}{e \times \angstrom^2}, respectively, and the Cl tip charge is modelled as a monopole with a charge of \SI{-0.3}{e}, where \si{e} is the elementary charge. The Lennard-Jones parameters for each atom type contained in our database of molecules are listed in Table \ref{tab:LJ}. To regularize the model and make it more robust, we randomize the tip-sample distance in the simulations within a \SI{0.5}{\angstrom} window (see Sec.\ "\nameref{sec:dist_rand}" above for more details). Additionally, the lateral equilibrium position of the probe particle is randomized within a disk of radius \SI{0.5}{\angstrom}.

\begin{table}[!hb]
    \centering
    \begin{tabular}{c c c}
        Element & $R_{ii}[\si{\angstrom}]$ & $E_{ii}[\si{eV}]$ \\
        \hline
        H & 1.4870 & 0.000681 \\
        C & 1.9080 & 0.003729 \\
        N & 1.7800 & 0.007372 \\
        O & 1.6612 & 0.009106 \\
        F & 1.7500 & 0.002645 \\
        Si & 1.9000 & 0.025490 \\
        P & 2.1000 & 0.008673 \\
        S & 2.0000 & 0.010841 \\
        Cl & 1.9480 & 0.011491 \\
        Br & 2.2200 & 0.013876 \\
        Xe & 2.1815 & 0.024344
    \end{tabular}
    \caption{Lennard-Jones parameters used in the probe particle simulations.}
    \label{tab:LJ}
\end{table}

\subsection*{Sensitivity of predictions to spring constant values}

\begin{figure*}[ht]
\centering
\includegraphics[width=130mm]{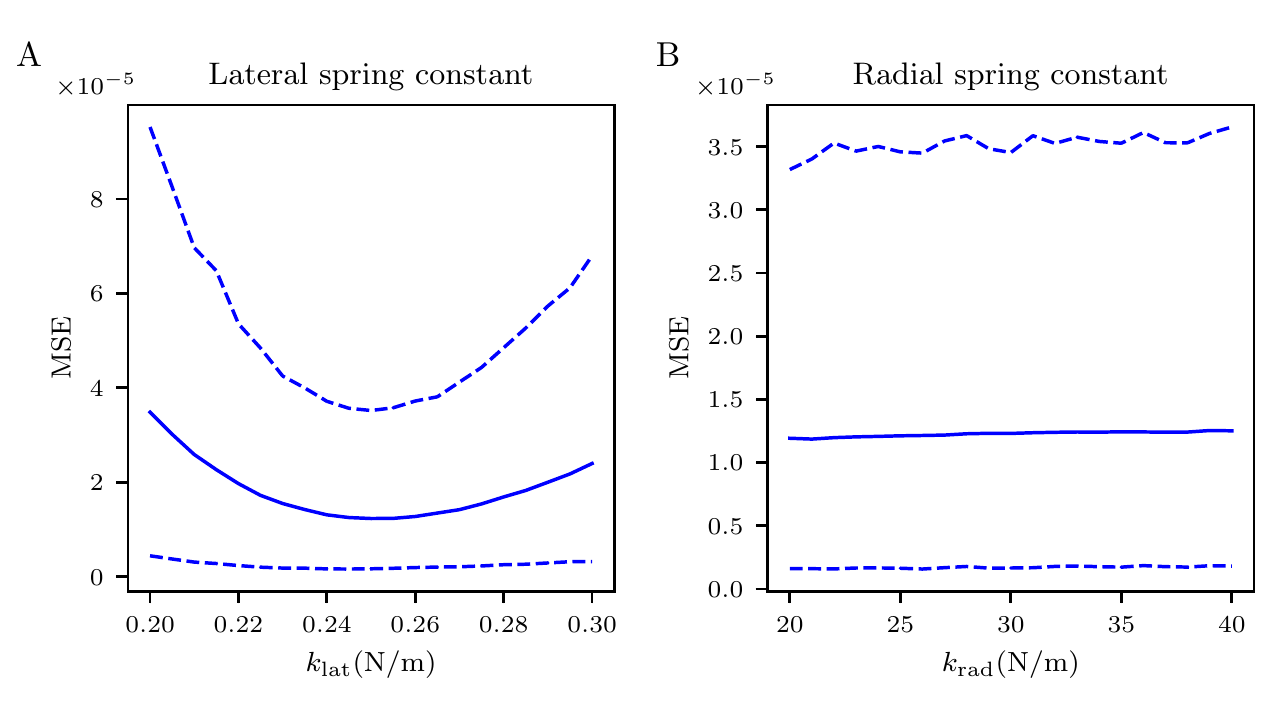}
\caption{Loss statistics for different values of the spring constants used in the AFM simulation for a subset of the test set. The plots show the MSE loss as a function of the value of (A) the lateral spring constant $k_{\mathrm{lat}}$ and (B) the radial spring constant $k_{\mathrm{rad}}$. The spring constants are altered for both the CO and the Xe tips. In both plots, the solid lines represent the mean loss, and the dashed lines represent the 5th and 95th percentiles, so that 90\% of the losses are contained within the region enclosed by the dashed lines.}
\label{stats_spring_constant}
\end{figure*}

We use fixed values for the lateral and radial spring constants in the AFM simulations in the training set. Since the tip condition can vary between AFM experiments, it is worth considering how sensitive the simulation and the predictions are to the chosen spring constant values. To this end, we run simulations on a subset of the test set varying the spring constant values in a range of $0.20 \dots 0.30~\si{N/m}$ for the lateral spring constant $k_{\mathrm{lat}}$, and $20 \dots 40~\si{N/m}$ for the radial spring constant $k_{\mathrm{rad}}$. On visual inspection of the simulated AFM images, for the radial spring constant there is no discernible difference between the different values in the chosen range. For the lateral spring constant, the differences are small, but can still be observed as a gradual change to a slightly sharper contrast in the close range with higher values of the spring constant. To quantify the sensitivity of the model predictions, we run the predictions for the simulated images and record the MSE loss as a function of the spring constant values (Fig.~\ref{stats_spring_constant}). The result is in line with the visual inspection of the AFM images: for the radial spring constant there is no significant difference in the loss values with different $k_{\mathrm{rad}}$ values, and for the lateral spring constant the loss increases smoothly when deviating from $k_{\mathrm{lat}}$ value used in the training. The loss increases more with decreasing $k_{\mathrm{lat}}$, reaching a value roughly 3 times the minimum loss at $k_{\mathrm{lat}} = \SI{0.25}{N/m}$. Keeping in mind that the MSE loss emphasizes outliers, this does still not correspond to a very large decrease in average performance. However, the result does show that the lateral spring constant is a parameter that could be worth randomizing during the training in the future to be more robust against changes in the tip condition.

\subsection*{Experimental}\label{sec:experimental}

\begin{figure*}[hp]
\centering
\includegraphics[width=160mm]{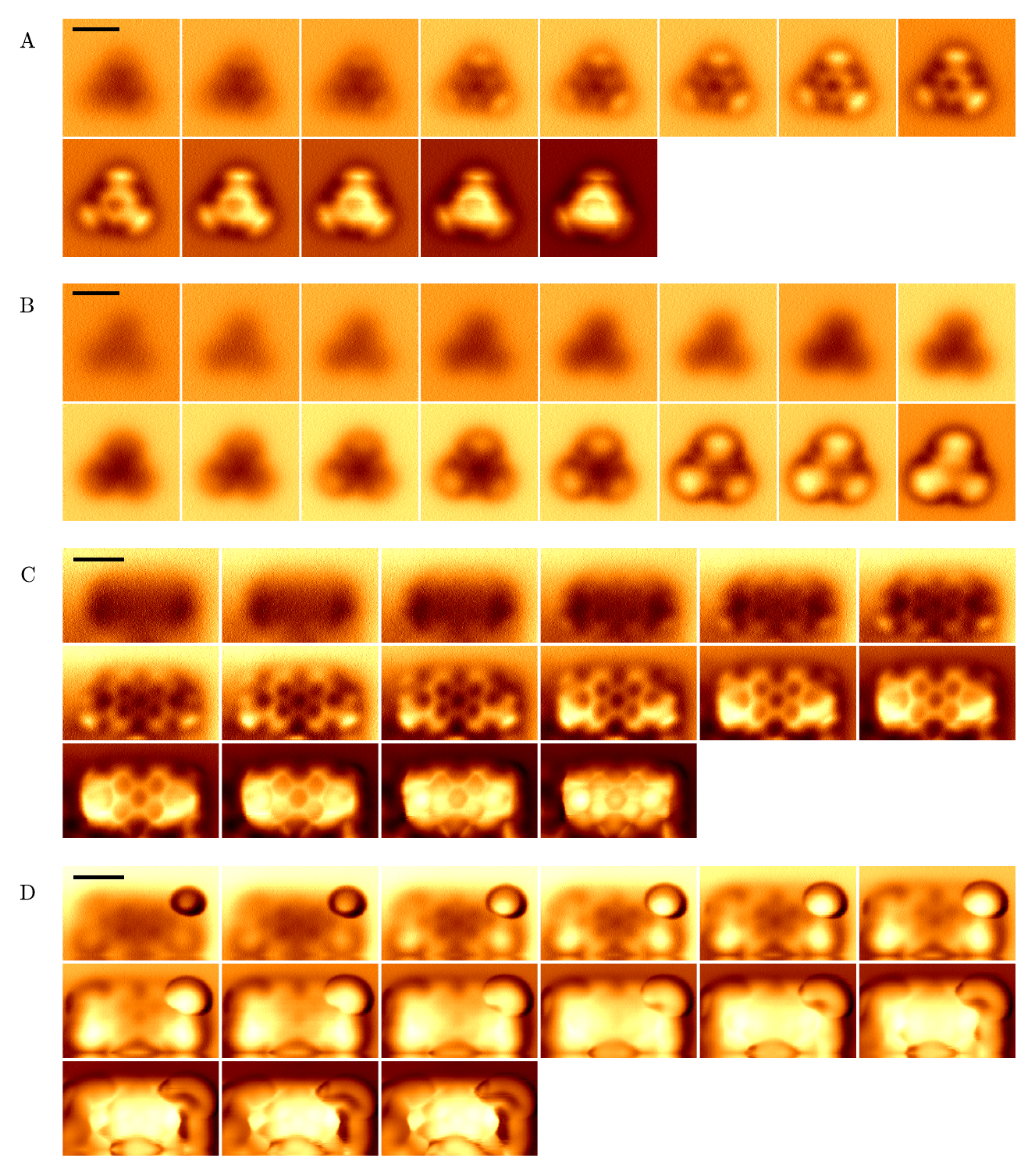}
\caption{Full sets of experimental AFM images for (A) BCB (CO), (B) BCB (Xe), (C) PTCDA (CO), and (D) PTCDA (Xe). The scale bars are \SI{5}{\angstrom} long.}
\label{afm_full}
\end{figure*}

\begin{figure*}[ht]
\ContinuedFloat
\centering
\includegraphics[width=160mm]{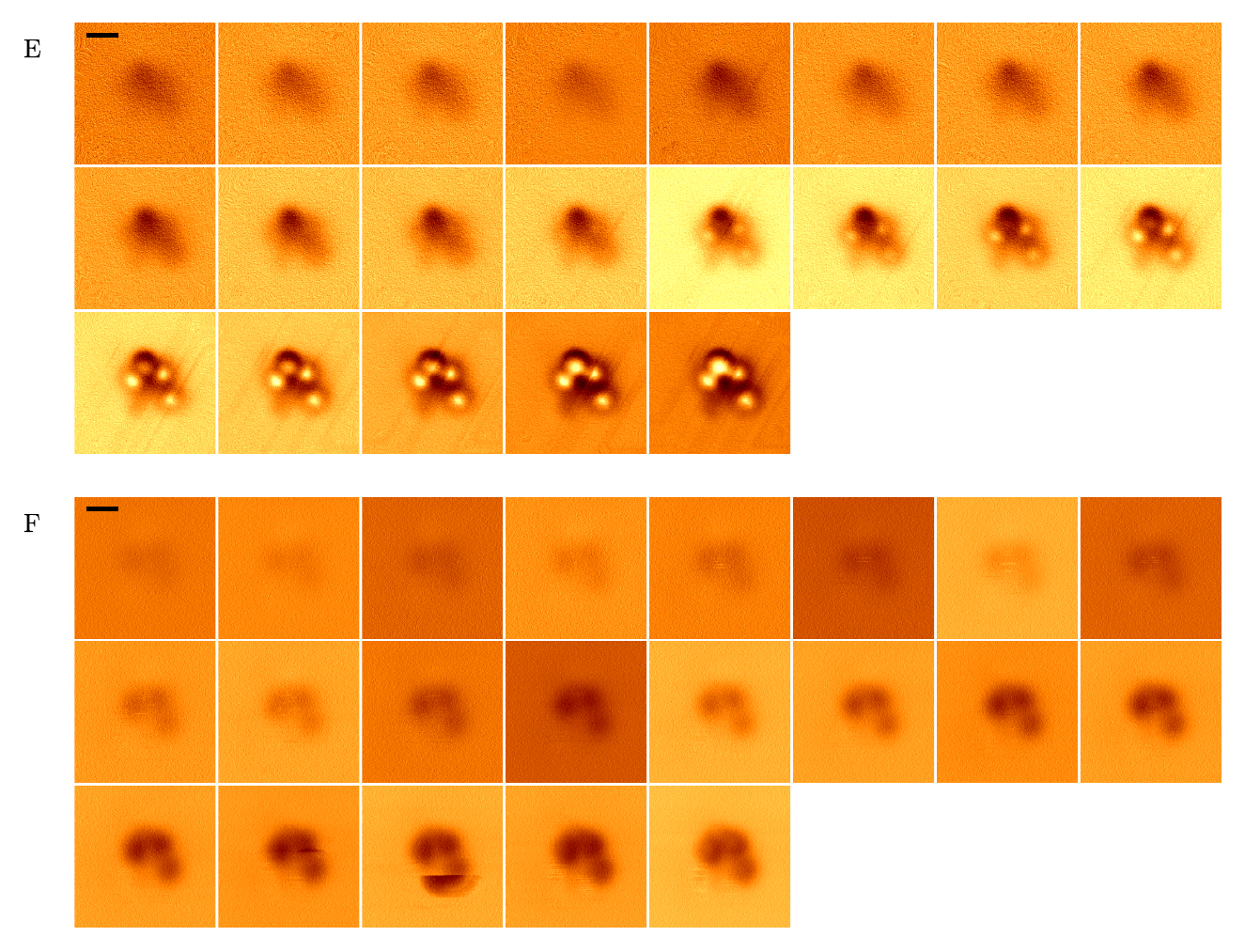}
\caption{(Continued) Full sets of experimental AFM images for (E) Water (CO), (F) Water (Xe). The scale bars are \SI{5}{\angstrom} long.}
\label{afm_full2}
\end{figure*}

The AFM images were taken on a combined non-contact AFM/STM system (CreaTec) with a commercial qPlus sensor with a Pt/Ir tip, operating at T $\approx$ 5K in ultrahigh vacuum at a pressure of \texttildelow 1 × 10\textsuperscript{-10} mbar. The qPlus sensor had a resonance frequency of \textit{f}\textsubscript{0} $\approx$ 30046 Hz, a quality factor \textit{Q} = 67714, and was always operating with an oscillation amplitude of A = 50 pm.

The Cu(111) substrate (MaTeck) was prepared by repeated Ne\textsuperscript{+} sputtering with a beam energy of 750 eV and ion current of 20 µA for 15 min followed by annealing at 520\texttildelow 550°C for 5 min. A flat Cu (111) surface with large terrace and minimum amount of impurities was often obtained within 3 cycles. The 1-Bromo-3,5-dichlorobenzene molecules (Sigma-Aldrich; purity 98\%) were deposited onto the substrate at \texttildelow 5 K through a variable leak valve \textbf{1} at a chamber pressure of 1 × 10\textsuperscript{-6} mbar for 30 seconds. Then the CO molecules (Praxair; purity 99.997\%) were deposited onto the substrate at ~5 K through a variable leak valve \textbf{2} at a chamber pressure of 1 × 10\textsuperscript{-6} mbar for less than 5 seconds. Finally, the Xe atoms (Fluka; purity 99.995\%) were deposited onto the substrate at ~5 K through the variable leak valve \textbf{1} at a chamber pressure of 1 × 10\textsuperscript{-6} mbar for 30 seconds.

Tip conditioning was usually performed by controlled contact with the Cu substrate and/or by applying a 1 second voltage pulse of 3\texttildelow10 V, both with feedback turned off. The tip was deemed as good when a symmetric contact mark was observed as well as a reasonably resolution of the organic molecules was achieved. The tip apex was believed to be covered with Cu atoms after these operations.

The constant height AFM images were taken with metal tips functionalized with a single CO molecule or a single Xe atom. The CO functionalization was achieved by applying a set-point of 8 mV /100 pA with the tip over a CO molecule, followed by turning off the feedback and then ramping the sample bias from zero to 2.6 V. A sudden decrease in the current happened at about 2.2 V indicates a successful functionalization. A subsequent scan over another CO showing sharp central protrusion can confirm the functionalization. After finishing with the CO tip, a bias ramping from zero to 3.6 V with feedback turned off can remove the CO while minimizing the perturbation to the structure of the metal tip apex. A sudden change of current at around 3.2 V often indicates a successful removal of the CO.

A second sequence of AFM images of the same molecule was taken with a Xe functionalized tip. The Xe functionalization was achieved by applying a set-point of 100 mV / 100 pA, followed by turning off the feedback and then bringing the tip into contact with a cluster of Xe atoms. A sudden decrease in current happened at \texttildelow3.5 Å advanced from the starting position indicates a successful transfer of a Xe atom. An STM scan with set-point of 100 mV / 100 pA capable of resolving individual Xe atoms inside the Xe cluster can further confirm such a functionalization.

The second experiment with PTCDA molecules (Sigma-Aldrich; purity 97\%) was done in a similar manner, except the PTCDA molecules were deposited onto the Cu(111) substrate at about 200 K using thermal sublimation.

The third experiment was done with water molecules (Sigma-Aldrich SKU38796; deionized). The water was purified before deposition, it was firstly boiled at 100°C to rid of any residual gas inside, and was then degassed thoroughly \emph{via} several freeze-pump-thaw cycles. During the experiment, water molecules were deposited \emph{via} a variable leak valve \textbf{3} aiming directly at the Cu(111) inside the scanner held at 5 K. The sample was subsequently heated up to 40 K, so that water molecules started to form clusters \cite{liriano_water_2017}. The sample was cooled back to 5 K thereafter. Xe and CO were deposited onto the surface with the same procedure as before.

\section*{Extended results}

\begin{figure*}[ht]
\centering
\includegraphics[width=150mm]{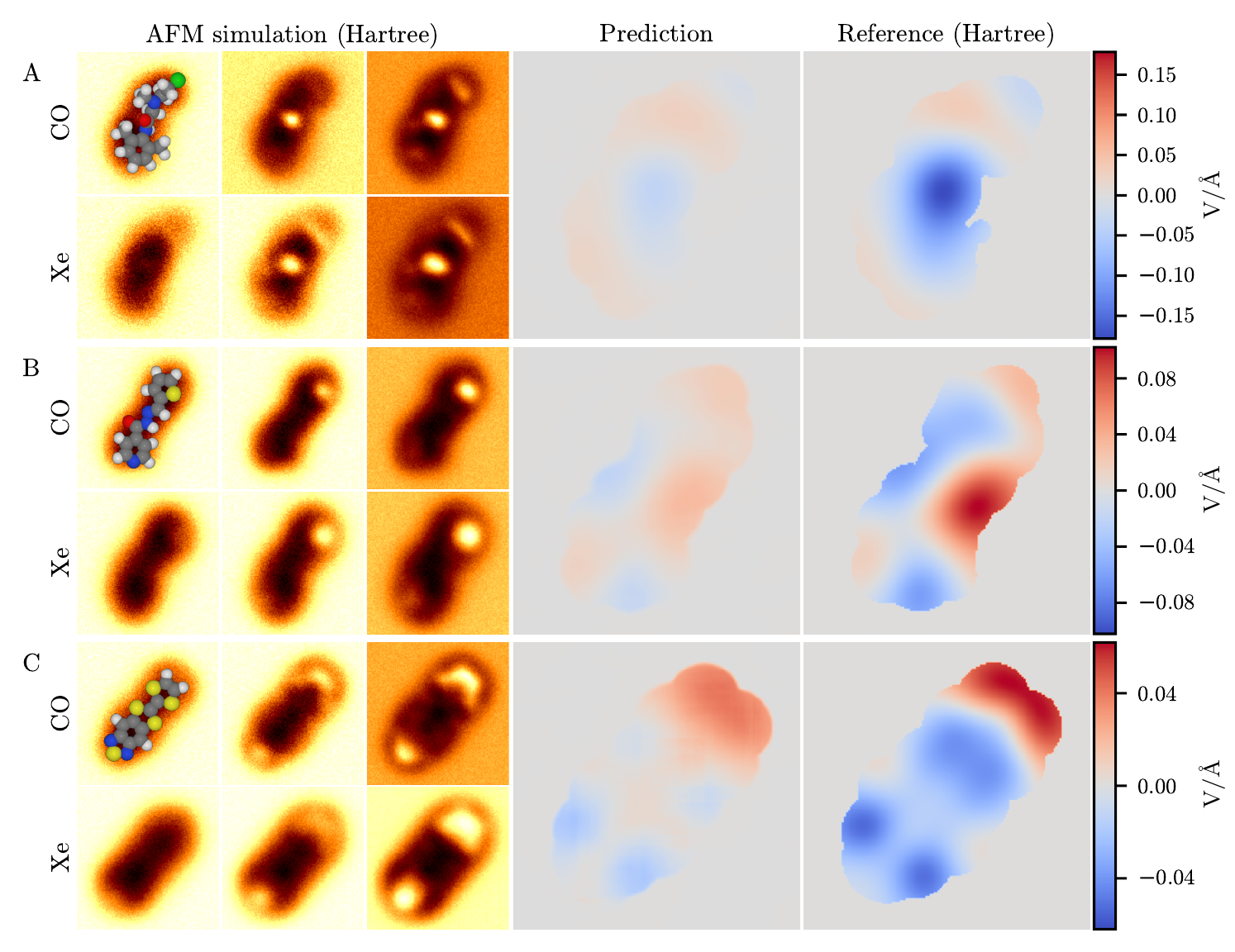}
\caption{Predictions for the benchmark examples using the DFT Hartree potential for electrostatics in the simulations. Compare to Fig.~2 in the main article.}
\label{sims_hartree}
\end{figure*}

\begin{figure*}[ht]
\centering
\includegraphics[width=150mm]{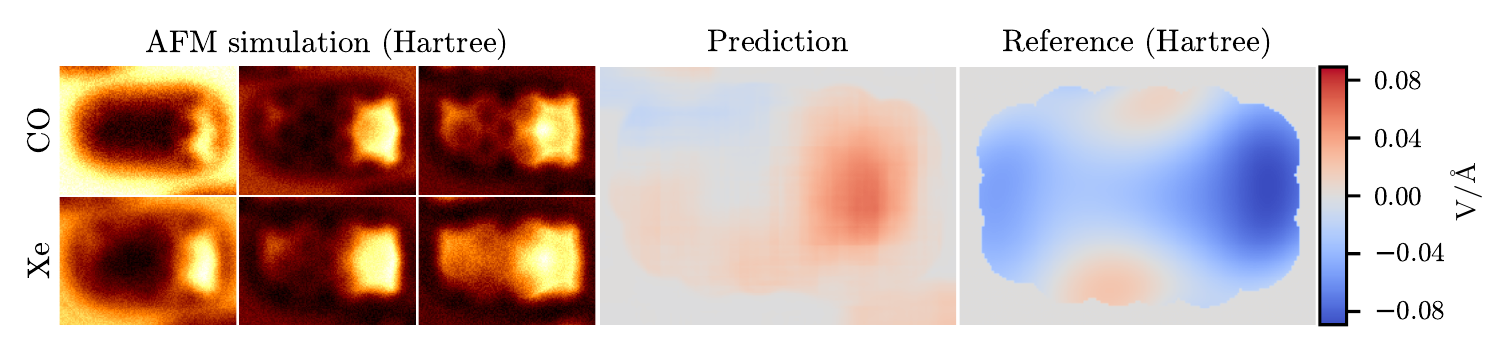}
\caption{Prediction and reference for on-surface geometry of PTCDA using the DFT Hartree potential for electrostatics in the AFM simulations and for the reference ES Map descriptor.}
\label{substrate}
\end{figure*}

\subsection*{Single-channel measurements}\label{sec:single-channel}

\begin{figure*}[ht]
\centering
\includegraphics[width=160mm]{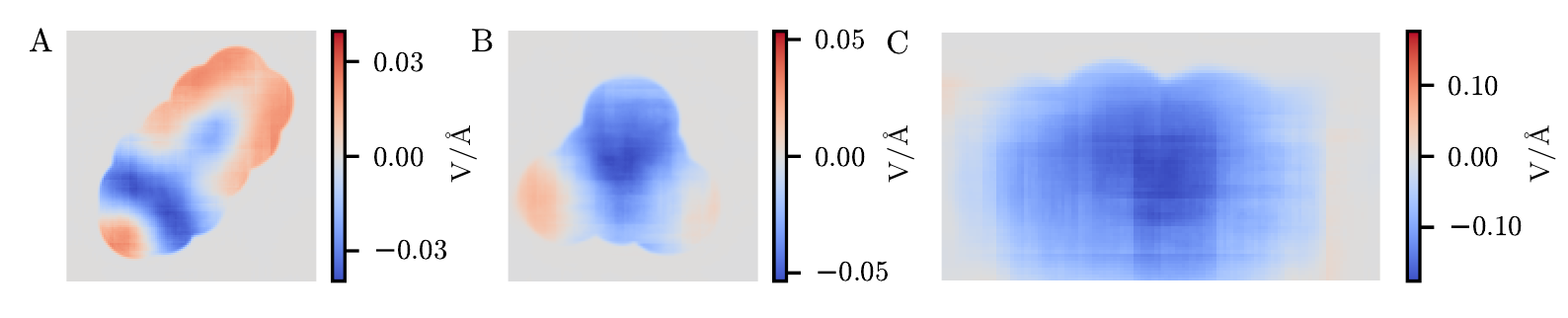}
\caption{Single-tip characterization of the benchmark examples in the main paper. Predictions are shown for (A) simulated data of TTF-TDZ and experimental data of (B) BCB and (C) PTCDA.}
\label{prediction_single}
\end{figure*}

Since the two-tip measurement presents an additional experimental challenge, we also try training a model using only a single-tip input of CO-AFM. This model is the same as the two-tip model, except that it lacks the other branch of layers in the beginning of the network. Fig.~\ref{prediction_single}A shows an example prediction on simulated data of the TTF-TDZ molecule. At a glance the prediction matches really well with the reference, but a closer inspection reveals that the magnitude of the field is not as accurate. The relative error for the prediction is $5.56\%$, more than twice the value for the two tip model. When measured on the whole test set, the average loss for the single-tip model is $77\%$ higher than for the two-tip model. Therefore, the single-tip model is less robust, but performance is not fatally worse.

We also apply the single-tip model to experimental data of BCB and PTCDA (Fig.~\ref{prediction_single}B,C) and find that the predictions are not very sensible. For BCB the model predicts mostly negative charge over the whole molecule with some positive regions over two of the halides. This does not match with either of the reference descriptors, where we expect to find the halides to be the most negative regions. The prediction for the PTCDA is similarly biased towards negative values in the middle of the molecule which in both reference descriptors is the least negative region. These results show that currently the addition of second channel of information is necessary to make the prediction work. Still, the relatively good performance on the simulations indicates that if the simulation model could be improved to be more accurate, then possibly even a single-channel measurement could be used for prediction.

\subsection*{Other tip combinations}\label{sec:other_tips}

\begin{figure*}[!ht]
\centering
\includegraphics[width=160mm]{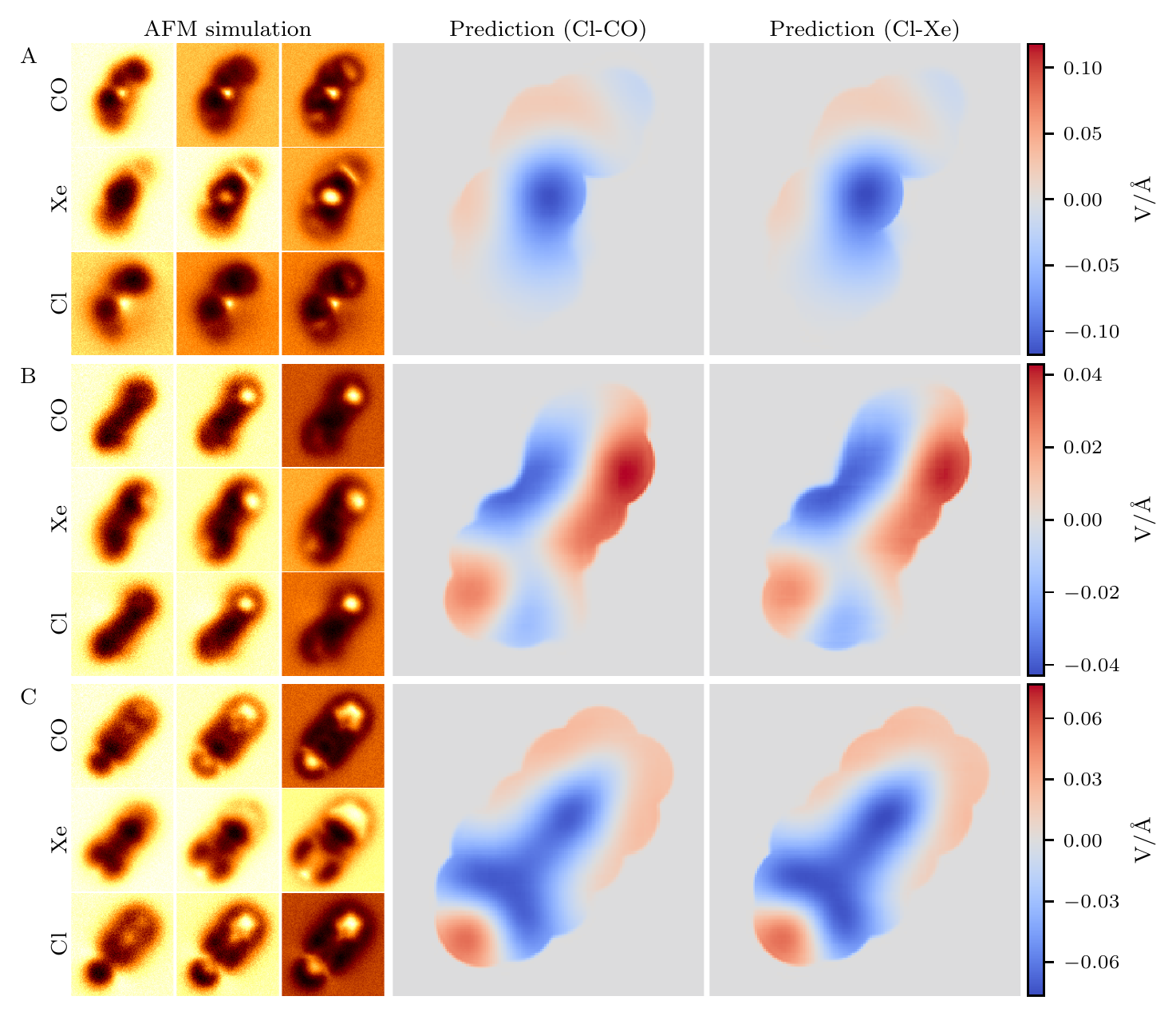}
\caption{Predictions with models trained on Cl-CO and Cl-Xe tip combinations on the benchmark examples. Compare to Fig.~2 in the main article.}
\label{sim_tips}
\end{figure*}

In order to show that the specific combination of CO and Xe tips is not special, we also generate simulations with the Cl tip and train models using the alternative tip combinations of Cl-CO and Cl-Xe. Figure \ref{sim_tips} shows example predictions for both tip combinations on simulations of the three benchmark examples introduced in the main article. On these examples, we find that the performance is roughly on par with the CO-Xe model, and for the losses on the test set we even find that the Cl-CO and Cl-Xe models have lower average losses than the CO-Xe model, by $46\%$ and $43\%$, respectively. In principle, any combination of tips can be used, as long as accurate simulated training data can be generated for them.

\subsection*{Distance dependence}\label{sec:distance}

\begin{figure*}[hp]
\centering
\includegraphics[width=150mm]{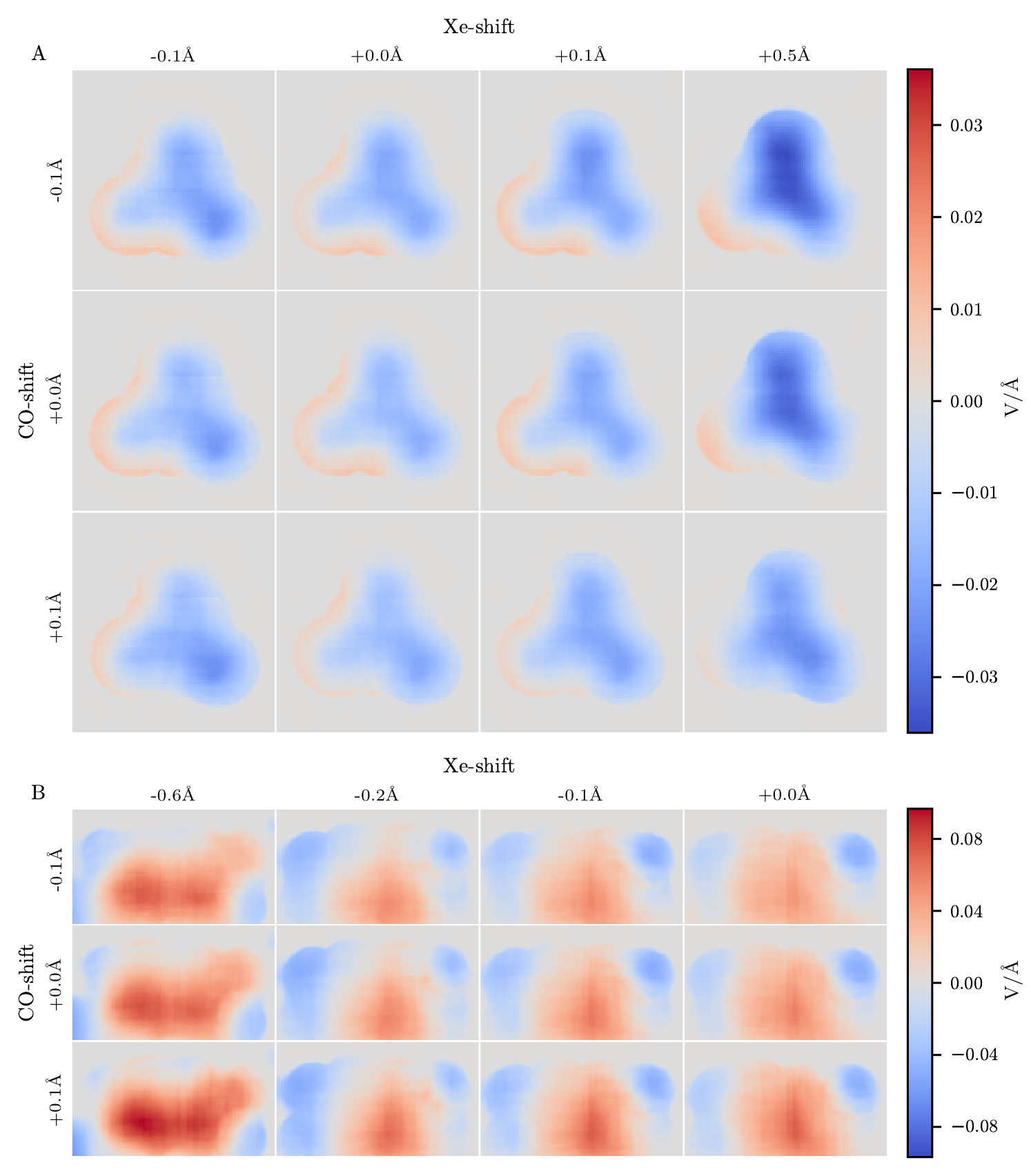}
\caption{Predictions at different tip-sample distances for experimental images of (A) BCB and (B) PTCDA. On each row the CO-AFM input has been shifted closer or further and on columns correspondingly the Xe-AFM input has been shifted. Here, the shift of +\SI{0.0}{\angstrom} corresponds to the distance used in the predictions in Figs.~3 and 4 of the main article. Negative shift corresponds to smaller tip-sample distance.}
\label{height_dependence}
\end{figure*}

The model is trained to take in AFM image stacks with six constant-height slices for both tips. In training the model, we consciously choose the tip-sample distance to be in range where the furthest images are mostly in the attractive regime, where only the overall shape of the molecule can be distinguished, and the closest images are in the repulsive regime, where at least some sharp atomic features are seen. However, we do not want to go too close to the molecule for two reasons. First, at close range the simulation data used for the training is less representative of the experimental case due to the simulation not taking into account any tip-induced relaxation of the sample. Second, at very close range the interaction between the tip and the sample is dominated by Pauli repulsion and the role of the electrostatics decreases.

In the experiments we have more than six slices for each measurement (see Figs.~\ref{afm_full} and \ref{afm_full2}), which leaves us with some room to choose which subset of images we use for the prediction. This selection process is still not automated and we have to use some judgment in choosing what is the best range for the data so that it best matches the training data, though we do augment the training with a variable range of distance in an attempt to be robust against variations in the distance. In Fig.~\ref{height_dependence} we explore for two of our experimental cases, BCB and PTCDA, what happens to the predictions when we vary the tip-sample distance of either the CO or the Xe tip. In both cases we find that small deviations within a range of $\SI{0.2}{\angstrom}$ do not change the predictions significantly. We also try larger deviations of +$\SI{0.5}{\angstrom}$ for BCB and $\SI{-0.6}{\angstrom}$ for PTCDA of the Xe tip to show that too large deviations start to alter the predictions. 

\subsection*{Possible extra electron in PTCDA}\label{sec:extra_electron}
\begin{figure*}[hb]
\centering
\includegraphics[width=100mm]{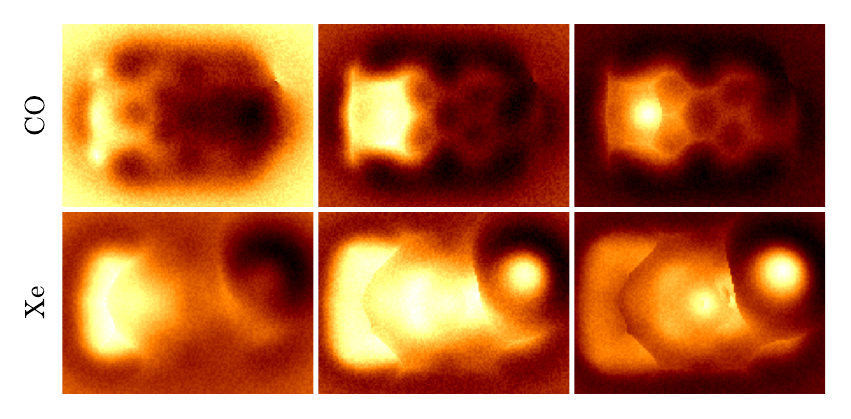}
\caption{AFM simulations of the on-surface PTCDA with one extra electron added to the top right oxygen. Note that the geometry has been flipped left-to-right compared to Fig.~\ref{substrate}A.}
\label{extra_electron}
\end{figure*}

In our experimental Xe-AFM images of PTCDA (Fig.~\ref{afm_full}D) we noted that there appears an unusual bright feature over the oxygen at the top right of the images. We speculate that this feature could be due to an extra electron trapped on that oxygen based on simulations with such an extra electron resulting in a similar bright feature, shown in Fig.~\ref{extra_electron}. In this simulation we took the on-surface geometry used for the DFT Hartree simulation in Fig.~\ref{substrate}A and did the simulation using Mulliken charges but adding an extra charge of \SI{-1}{e} to the oxygen at the top right of the molecule. The result is a large bright spot surrounded by a dark halo in the Xe-AFM image, somewhat similar to the one in the experimental images. Furthermore, in the CO-AFM simulation there appears a sharp change in contrast over the same oxygen, which on the closer distance makes the oxygen appear very dim compared to the rest of the molecule, which also bears similarity to the experimental images (Fig.~\ref{afm_full}C). It is, however, unclear if such extra electron would stay trapped on the oxygen and not transfer to the substrate or tip even when the molecule is being pushed by the AFM probe.

\subsection*{Surface tilt effect on model predictions}\label{sec:surface_tilt}
\begin{figure*}[ht]
\centering
\includegraphics[width=150mm]{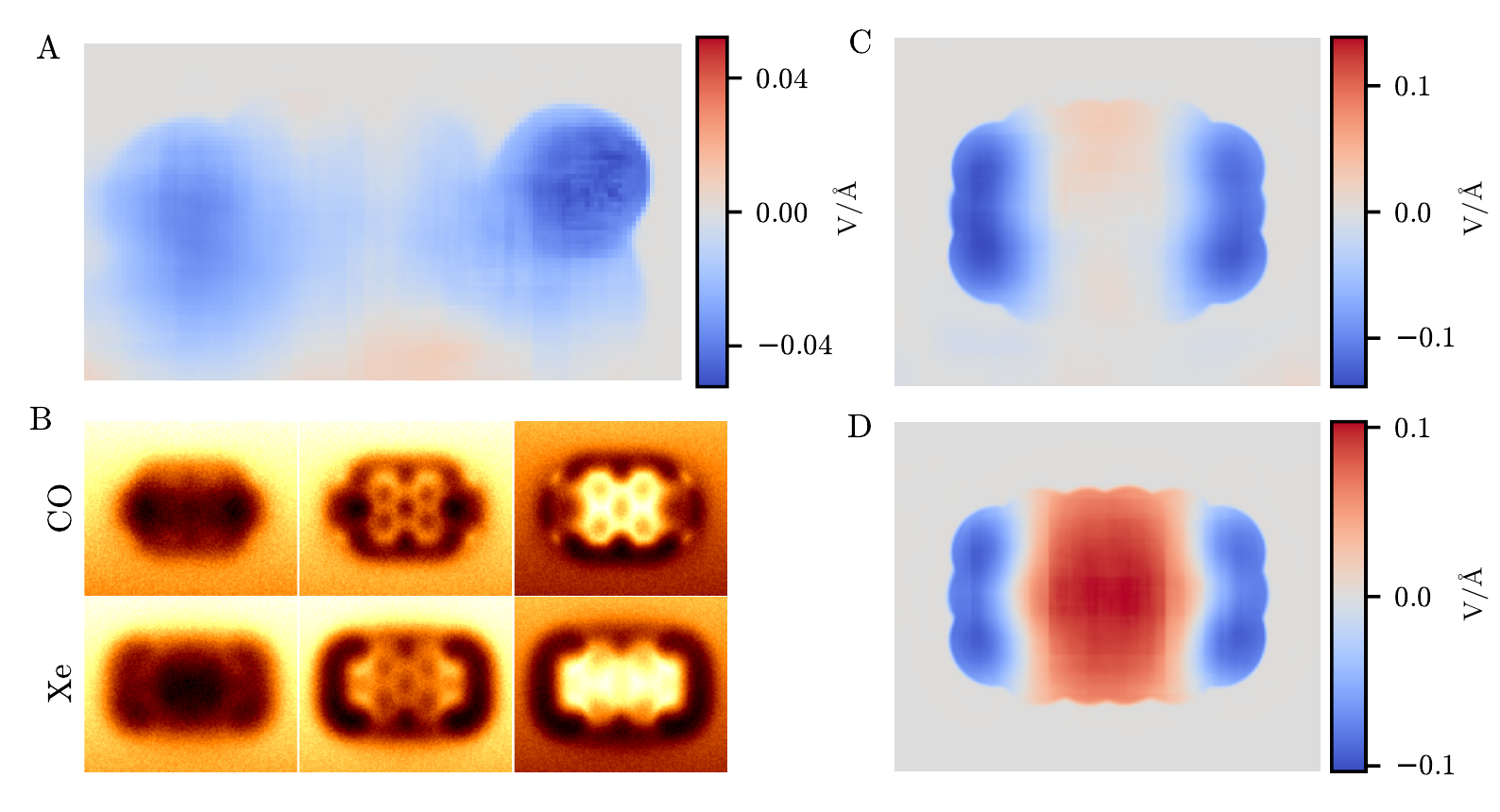}
\caption{Surface tilt effect on model predictions. (A) ES Map prediction of experimental AFM images of PTCDA on a model trained without background gradient augmentation. (B) Simulated AFM images of PTCDA with added background gradient. Using the input data in (B) we predict the ES Map on models trained (C) without and (D) with the background gradient augmentation.}
\label{surface_tilt}
\end{figure*}

The PTCDA dataset presented us with the additional challenge that the experiment was performed at a slight tilt which resulted in a gradient in the background of the image. When we artificially added a similar gradient to the simulated AFM images of PTCDA (Fig.~\ref{surface_tilt}B), the prediction (Fig.~\ref{surface_tilt}C) failed by incorrectly predicting the positive region in the middle of the molecule as being close to neutral. The prediction of the experimental data (Fig.~\ref{surface_tilt}A) showed a similar pattern.

Motivated by this finding, we augmented the training of the model with these background gradients, implemented by adding a plane with a set gradient to each AFM image set. The direction of the gradient is uniform random, and the magnitude of the gradient is randomized such that the range of values in the plane is at most 30\% of the range of values in the image set. The zero-point of the plane is always at the center of the images. With this augmentation, we found that the predictions improved both on the simulated images (Fig.~\ref{surface_tilt}D) and the experimental prediction also became more consistent. Being robust against these kinds of tilts could be useful in situations where a tilted planar section of a molecule needs to be characterized \cite{albrecht_characterization_2015}.

\section*{References}
\printbibliography[heading=none]
